\documentclass{JHEP3}
\usepackage{amsmath,amsfonts,subfigure}%,graphicx,epsfig}
\usepackage{graphicx}
\DeclareMathAlphabet   {\mathsc}{OT1}{cmr}{m}{sc}

\def\[{\left [}
\def\]{\right ]}
\def\({\left (}
\def\){\right )}

\newcommand{\lbr}{\left\{}
\newcommand{\rbr}{\right\}}
\newcommand{\beq}{\begin{equation}}
\newcommand{\eeq}{\end{equation}}
\newcommand{\bea}{\begin{eqnarray}}
\newcommand{\eea}{\end{eqnarray}}
\newcommand{\wtd}[1]{\widetilde{#1}}
\newcommand{\GeV}      {~\mathrm{GeV}}
\newcommand{\TeV}      {~\mathrm{TeV}}

\newcommand{\SUSY}     {\mathsc{susy}}

\newcommand{\order}{\mathcal{O}}

\newcommand{\gappeq}{\mathrel{\rlap {\raise.5ex\hbox{$>$}}
{\lower.5ex\hbox{$\sim$}}}}
\newcommand{\lappeq}{\mathrel{\rlap{\raise.5ex\hbox{$<$}}
{\lower.5ex\hbox{$\sim$}}}}

\newcommand{\met}{\not{\hspace{-.05in}{E_T}}}
\title{Phenomenological Implications of Deflected Mirage Mediation:  Comparison with Mirage Mediation}
\author{Baris~Altunkaynak and  Brent~D.~Nelson  \\
Department of Physics, Northeastern University,
Boston, MA, 02115 USA}
\author{Lisa L.~Everett, Ian-Woo~Kim, and Yongyan~Rao\\
Department of Physics, University of Wisconsin,
Madison, WI 53706, USA}

\preprint{MADPH-10-1553}

\abstract{We compare the collider phenomenology of mirage mediation and deflected mirage mediation, which are two recently proposed ``mixed" supersymmetry breaking scenarios motivated from string compactifications.  The scenarios differ in that deflected mirage mediation includes contributions from gauge mediation in addition to the contributions from gravity mediation and anomaly mediation also present in mirage mediation.  The threshold effects from gauge mediation can drastically alter the low energy spectrum from that of pure mirage mediation models, resulting in some cases in a squeezed gaugino spectrum and a gluino that is much lighter than other colored superpartners.  We provide several benchmark deflected mirage mediation models and construct model lines as a function of the gauge mediation contributions, and discuss their discovery potential at the LHC.}

\begin{document}

%\DeclareGraphicsExtensions{.pdf,.gif,.jpg}

\maketitle

%---------------------
\section{Introduction}
\label{sec:intro}

TeV scale softly broken supersymmetry (SUSY) (for recent reviews, see \cite{Martin:1997ns,Chung:2003fi}) is one of the best-motivated candidates for physics beyond the Standard Model (SM).  Theories with softly broken supersymmetry, such as the minimal supersymmetric standard model (MSSM), will be tested thoroughly at the Large Hadron Collider (LHC).  The collider phenomenology of such theories depends in detail on the associated soft supersymmetry breaking terms, which in turn are largely governed by the way in which supersymmetry breaking is mediated from a hidden sector to the SM fields.  As the number of possible soft supersymmetry breaking parameters is vast even in minimal extensions of the SM such as the MSSM, it is fruitful to develop models of supersymmetry breaking.  Exploring the possible mediation mechanisms for supersymmetry breaking is crucially important for LHC tests of the hypothesis that supersymmetry is present at TeV energy scales.

Viable supersymmetry breaking models are based on the hidden sector paradigm, in which supersymmetry is broken in a hidden or secluded sector and communicated to the observable sector via the interactions of mediator fields.  As is well known, the phenomenological implications of these models are largely insensitive to the details of the hidden sector, and instead are governed by the mediation mechanism that is responsible for the transmission of supersymmetry breaking to the MSSM fields.
Most models of supersymmetry breaking involve one of the three most popular mediation mechanisms: (i) gravity mediation, (ii) gauge mediation, and (iii) (braneworld-motivated) ``bulk" mediation models.  Gravity-mediated terms \cite{grav1} arise from couplings that vanish as the Planck mass $M_P\rightarrow \infty$; examples include the (tree-level) minimal supergravity (mSUGRA) model (also known as the constrained MSSM (CMSSM)) and modulus mediation models  \cite{modulus}, and the (loop-suppressed) anomaly mediation models \cite{anomaly}.  Gauge-mediated terms arise from loop diagrams involving new messenger fields with SM charges \cite{gauge1,gauge2,Giudice:1998bp}.   Bulk-mediated terms arise from bulk mediator fields in braneworld scenarios.  Examples include gaugino mediation \cite {gaugino} and $Z'$ mediation \cite{zprime}.  Certain gravity-mediated models, such as the pure anomaly mediation scenario \cite{anomaly} (which requires sequestering), are also bulk mediation models.    
Typically, one of these mediation mechanisms is assumed to dominate (see e.g.~\cite{Allanach:2002nj}), for simplicity and practicality and/or to solve a given problem of the MSSM, such as the $\mu$ problem or the flavor/CP problems.   

A complementary approach is to consider models in which more than one mediation mechanism plays an important role.  Such ``mixed" scenarios are motivated within string-motivated constructions, such as the Kachru-Kallosh-Linde-Trivedi (KKLT) approach to moduli stabilization \cite{Kachru:2003aw}. A prototype KKLT-motivated example is {\it mirage mediation}~\cite{Choi:2004sx,Choi:2005ge}, a phenomenological model in which the tree-level gravity (modulus)-mediated terms and the (loop-suppressed) anomaly-mediated terms are comparable in size, contrary to naive expectations.  As a result, the soft masses unify at a scale that is typically well below the scale where they are generated, resulting in ``mirage" unification.   Mirage mediation has distinctive phenomenological implications compared with standard minimal supergravity models \cite{Choi:2005uz,miragepheno,Choi:2006bh,Choi:2008hn}.  These features include a gaugino mass pattern that is typically more squeezed than the standard mSUGRA/CMSSM gaugino mass pattern \cite{Choi:2007ka}  and reduced low energy fine-tuning \cite{littlehierarchy}, for which the details depend on the ratio of the gravity-mediated contributions to the anomaly-mediated contributions. 

The recently proposed {\it deflected mirage mediation} scenario is an extension of mirage mediation in which gauge-mediated supersymmetry breaking terms are also present and competitive in size to the gravity-mediated and anomaly-mediated soft terms \cite{Everett:2008qy,Everett:2008ey}.   This framework is denoted as deflected mirage mediation because the mirage unification scale in the gaugino sector is shifted or ``deflected" from its mirage mediation value due to threshold effects associated with the gauge mediation messengers fields. In deflected mirage mediation, the gauge-mediated terms arise from the couplings of an additional matter modulus field $X$ and vectorlike messenger pairs with nontrivial SM quantum numbers.  The ratio of the gauge-mediated and anomaly-mediated terms depends on the details of the stabilization mechanism for the mediator field $X$.  In \cite{Everett:2008ey}, it was found that if the stabilization mechanism is dominated by supersymmetry breaking effects, generically the gauge-mediated and anomaly-mediated contributions are comparable.  Deflected mirage mediation provides a general framework in which to explore mixed supersymmetry breaking scenarios at the LHC, where well-known single mediation mechanism models can be recovered by judiciously adjusting dimensionless parameters in the theory.

In this paper, we explore the question of how the collider phenomenology of deflected mirage mediation differs from that of standard mirage mediation, following previous work on the sparticle spectrum \cite{Everett:2008ey,Choi:2009jn} and dark matter constraints \cite{Holmes:2009mx}.  More precisely, we focus our attention on the effects of the gauge mediated terms by varying the ratio of gauge to anomaly mediation contributions and/or the number of messenger pairs, as compared to benchmark KKLT mirage mediation scenarios.   The differences in the collider phenomenology of the deflected mirage mediation scenarios compared to the corresponding pure mirage mediation scenarios depend primarily on the deflection of the gaugino mass mirage unification scale.  This depends in turn on the size of the messenger scale and whether the threshold effects are large or small.  For small threshold effects, the pattern of soft masses does not differ greatly from the corresponding pure mirage mediation limit.  However, if the threshold effects are large, one can have situations in which the gaugino mirage unification scale can be deflected from a high scale value to the TeV scale.  In such situations, the gaugino mass spectrum is squeezed, resulting in gluinos that are typically lighter than other colored superpartners. The LHC phenomenology is thus dominated by gluino production in this case, with soft decay products due to the compressed chargino and neutralino mass spectrum.

The outline of this paper is as follows.  In Section~\ref{sec:theory}, we review the theoretical framework of deflected mirage mediation and enumerate the supersymmetry breaking terms of the theory.  
An overview of mirage unification and the properties of the resulting low energy mass spectra in both mirage mediation and deflected mirage mediation is given in Section~\ref{sec:scan}.  We then discuss the collider phenomenology of benchmark deflected mirage mediation models and compare it to that of pure mirage mediation benchmark scenarios in Section~\ref{sec:collider}.   In Section~\ref{sec:conc}, we provide our conclusions and outlook.	
	
%---------------------
\section{Theoretical Framework}
\label{sec:theory}
 
We begin with a brief review of mirage mediation, a phenomenological model motivated from the KKLT flux compactification approach within Type IIB string theory \cite{Kachru:2003aw}.  In this setup, the MSSM fields are confined to a stack of D branes that are localized in a higher-dimensional bulk Calabi-Yau space, and the hidden (supersymmetry breaking) sector consists of anti-branes at the tip of the warped throat geometry.  The tree-level gravity mediation terms are 
\begin{equation}
\label{graveq}
m_{\rm soft}^{({\rm modulus})} \sim  \frac{F^T}{T+\overline{T}},
\end{equation}
where  $T$ is the K\"{a}hler modulus.  The anomaly mediation terms are
\begin{equation}
m_{\rm soft}^{({\rm anomaly})}\sim \frac{1}{16\pi^2} \frac{F^C}{C},
\end{equation}
where $C$ is the conformal compensator of the gravity multiplet.  In the KKLT scenario, there is a cancellation between the superpotential terms from the fluxes and the nonperturbative terms, leading to a supersymmetry-preserving vacuum with stabilized moduli but with a negative cosmological constant.  Supersymmetry is then broken by an uplifting potential of the form $(T+\overline{T})^{-n_p}$, where $n_p=2$ in the KKLT model, due to the anti-branes at the tip of the warped throat.  After cancelling the cosmological constant, the following mirage mediation relation is obtained \cite{Choi:2004sx,Choi:2005ge}:
\begin{equation}
 \frac{F^T}{T+\overline{T}}\sim \frac{1}{\ln(M_P/m_{3/2})} \frac{F^C}{C},
\end{equation}
in which $M_P  = 2.4 \times 10^{18} \mbox{ GeV}$ is the reduced Planck mass, and $m_{3/2}$ is the gravitino mass.  As $m_{3/2}$ is typically $\sim 100$ TeV in this class of models, the factor of $\ln(M_P/m_{3/2})$ is numerically close to $4\pi^2$. Hence, the tree-level gravity mediation terms are comparable to the anomaly mediation terms in mirage mediation.
 
In deflected mirage mediation, the observable sector matter content also includes a gauge singlet $X$ and vectorlike messenger pairs $\Psi$, $\overline{\Psi}$ with SM gauge charges, which are fields that are generically present in string-derived models.  In general, $X$ can acquire an F term vacuum expectation value, leading to gauge mediated terms of the form
\begin{equation}
m_{\rm soft}^{({\rm gauge})}\sim \frac{1}{16\pi^2} \frac{F^X}{X}.
\end{equation}
Depending on the stabilization mechanism for $X$, it was shown  in \cite{Everett:2008qy,Everett:2008ey} that in general,
\begin{equation}
\frac{F^X}{X}\sim \frac{F^C}{C}, 
\end{equation}
such that $m_{\rm soft}^{({\rm gauge})}$ is comparable to $m_{\rm soft}^{({\rm anom})}$ and $m_{\rm soft}^{({\rm grav})}$ in deflected mirage mediation. 

To see this more clearly, let us begin with the effective supergravity theory of KKLT-inspired models.  Labeling the observable sector (MSSM) fields as $\Phi$ and taking a diagonal matter metric for simplicity, the K\"{a}hler potential at leading order is 
\begin{equation}
\label{fullkahler}
K = - 3 \log (T+\overline{T})  + \frac{ X \overline X}{(T+\overline{T})^{n_X}} + \frac{ \Phi_i \overline{\Phi}_i}{(T+\overline{T})^{n_i}}, 
\end{equation}
where $n_X$ and $n_i$ are the modular weights of $X$ and $\Phi_i$, respectively.  The superpotential is
\begin{equation}
\label{superpot}
W=W_0+W_1(X)+\lambda X\Psi\overline \Psi +W_{\rm MSSM}.
\end{equation}
In Eq.~(\ref{superpot}), $W_0$ is the part of the superpotential that governs (together with the uplifting potential) the supersymmetry breaking effects \cite{Kachru:2003aw}, $W_1(X)=\lambda_n X^n$ describes the possible self-couplings of $X$, and $W_{\rm MSSM}$ takes the form
\begin{equation}
W_{\rm MSSM}=\mu^0_{ij} \Phi_i\Phi_j +y^0_{ijk}\Phi_i\Phi_j\Phi_k,
\label{MSSMW}
\end{equation}
in which $\mu^0_{ij}$ are supersymmetric mass parameters, and $y^0_{ijk}$ are the (unnormalized) Yukawa couplings.  The gauge kinetic functions are assumed to take the form 
\begin{eqnarray}
f_a (M_{\rm G}) = T. 
\end{eqnarray}
The messengers  are taken to be $\Psi$, $\overline{\Psi}$ are 5, $\bar{5}$ representations of $SU(5)$, as is standard in many models of gauge mediation.  Here $N$ will denote the number of such messenger pairs.

Upon computing the soft terms using standard supergravity techniques,  we obtain the observable sector soft supersymmetry breaking Lagrangian, which is of the usual form
\begin{eqnarray}
\mathcal{L}_{\rm soft}=-m^2_i |\Phi^i|^2-\left [ \frac{1}{2} M_a \lambda^a \lambda^a + A_{ijk} y_{ijk} \Phi^i \Phi^j \Phi^k +\mbox{h.c.} \right ],
\end{eqnarray} 
in which $m^2_i$ are the soft scalar mass-squared parameters, $M_a$ are the gaugino masses, and $A_{ijk}$ are trilinear scalar interaction parameters.\footnote{These terms are defined in the field basis in which the kinetic terms are canonically normalized; the physical Yukawa couplings $y_{ijk} = y^0_{ijk} / (Z_i Z_j Z_k)^{1/2}$ are in the 
definition of trilinear terms.}    

Recalling that above the mass scale of the messengers $M_{\rm mess}\equiv \lambda \langle X \rangle$, the beta functions depend on not only the MSSM fields, but also on the messenger pairs,  
the soft terms at the GUT scale $M_{\rm G}$ and the messenger threshold effects at $M_{\rm mess}$ are as follows:\\

\noindent {\bf Gaugino Masses.} The gaugino mass parameters are given by
\begin{eqnarray}
  M_a (M_{\rm G}) &=& \frac{F^T}{T+\overline{T}}
  + \frac{g_0^2}{16\pi^2}  b'_a\frac{F^C}{C} \\ \label{softgaugino}
  M_a (M_{\rm mess}^-)& =& M_a(M_{\rm mess}^+)
+ \Delta M_a,
\end{eqnarray}
in which the threshold corrections are
\begin{eqnarray}
  \Delta M_a = - N\frac{g_a^2(M_{\rm mess})}{16\pi^2} \left( \frac{F^C}{C}
  + \frac{F^X}{X} \right).
   \label{softgauginothresh}
\end{eqnarray}
Here $g_0$ is the unified gauge coupling at $M_{\rm G}$, and the beta functions $b^\prime_a$ are related to their MSSM counterparts by 
$b^\prime_a = b_a + N$,
with  
$(b_3, b_2, b_1) = (-3, 1, \frac{33}{5})$ 
(in our conventions, $b_a<0$ for asymptotically free theories).\\

\noindent {\bf  Trilinear terms.}
The trilinear terms are $A_{ijk} = A_i+ A_j + A_k$, with
\begin{eqnarray}
A_i (\mu = M_{\rm G}) &=& \left (1-n_i \right) \frac{F^T}{T+\overline{T}}
- \frac{\gamma_i}{16\pi^2} \frac{F^C}{C},  \label{softA}
\end{eqnarray}
where $\gamma_i$ is the anomalous dimension of $\Phi_i$. \\

\noindent {\bf Soft scalar masses. }
The scalar mass-squared parameters are given by
\begin{eqnarray}
m_i^2 (\mu = M_{\rm G}) = \left (1-n_i \right) \left|\frac{F^T}{T+\overline{T}}\right|^2
%\nonumber \\
%&& 
-\frac{\theta'_i}{32 \pi^2} \left( \frac{F^T}{T+\overline{T}}\frac{F^{\overline{C}}}{\overline{C}} + \mbox{h.c.} \right) \nonumber \
\label{softscalar}
 - \frac{\dot{\gamma}'_i}{(16\pi^2)^2} \left|\frac{F^C}{C}\right|^2,
%\nonumber \\
\end{eqnarray}
\begin{eqnarray}
m_i^2 (\mu = M_{\rm mess}^-) &=& m_i^2 (\mu = M_{\rm mess}^+)
+ \Delta m_i^2,
\end{eqnarray}
where the threshold corrections are
\begin{eqnarray}
\Delta m_i^2 & = &\sum_a 2 c_a N
\frac{g_a^4 (M_{\rm mess}) }{(16\pi^2)^2}
 \left( \left|\frac{F^X}{X}\right|^2 + \left|\frac{F^C}{C}
\right|^2
+ \frac{F^X}{X}
 \frac{F^{\overline{C}}}{\overline{C}} +\mbox{h.c.} \right). \label{softscalarthresh}
\end{eqnarray}
In the above, $c_a$ is the quadratic Casimir, and $\gamma_i$, $\dot{\gamma}_i$, $\theta_i$ ($\gamma_i'$, $\dot{\gamma}_i'$, $\theta_i'$) are listed in Appendix A.
%\end{itemize}
We now replace the F terms with the parameterization given in  \cite{Everett:2008qy,Everett:2008ey}, as follows:
\begin{eqnarray}
\label{param1}
\frac{F^C}{C}&=&\alpha_m\ln \frac{M_P}{m_{3/2}} \frac{F^T}{T+\overline{T}}= \alpha_m\ln \frac{M_P}{m_{3/2}} M_0 \\
\label{param2}
\frac{F^X}{X}&=&\alpha_g \frac{F^C}{C}=\alpha_g\alpha_m \ln\frac{M_P}{m_{3/2}} M_0,
\end{eqnarray}
in which $M_0 \equiv F^T/(T+\overline{T}) $ sets the overall scale of the soft terms.  The dimensionless parameter $\alpha_m$ is the $\alpha$ parameter of mirage mediation: it denotes the relative importance of anomaly mediation with respect to gravity mediation.  In the KKLT case $\alpha_m=1$; the case of $\alpha_m=2$ ({\it i.e}, an uplifting potential with $n_p=1$, which has no obvious string theory realization) has also been considered in the literature, for reasons that will be clear shortly.  The dimensionless parameter $\alpha_g$ denotes the relative importance of the gauge-mediated terms with respect to the anomaly-mediated terms.  The values of $\alpha_g$ depend on the details of the stabilization of $X$ \cite{Everett:2008qy,Everett:2008ey}. For radiative stabilization, 
\begin{equation}
\frac{F^X}{X}=-\frac{F^C}{C},
\end{equation}
while for self-couplings of the form $W_1=\lambda_nX^n$, 
\begin{equation}
\frac{F^X}{X}=-\frac{2}{n-1}\frac{F^C}{C}.
\end{equation}
Here $n$ is restricted to the values $n\geq 3$  or $n<0$, {\it i.e.}, stabilization by higher-dimensional operators or non-perturbative effects, respectively.  With the parametrization given in Eqs.~(\ref{param1})--(\ref{param2}), the soft terms at $M_G$ take the form
\begin{eqnarray}
\label{gaugino1}
M_a(M_{\rm G})&=&M_0\left [1+\frac{g_0^2}{16\pi^2}b_a^\prime \alpha_m \ln \frac{M_P}{m_{3/2}}\right ] ,\\
\label{trilinear1}
A_i(M_{\rm G})&=&M_0 \left [(1-n_i)-\frac{\gamma_i}{16\pi^2} \alpha_m \ln \frac{M_P}{m_{3/2}}\right ] ,\\
\label{mssq1}
m_i^2(M_{\rm G})&=&M_0 ^2 \left [(1-n_i)-\frac{\theta'_i}{16 \pi^2} \alpha_m \ln \frac{M_P}{m_{3/2}} -\frac{\dot{\gamma}'_i}{(16\pi^2)^2}\left (\alpha_m \ln \frac{M_P}{m_{3/2}}\right )^2 \right ],
\end{eqnarray}
and the threshold terms are given by
\begin{eqnarray}
\label{threshgaug1}
\Delta M_a&=& -M_0 N\frac{g_a^2(M_{\rm mess})}{16\pi^2}   \alpha_m \left (1 +\alpha_g \right ) \ln \frac{M_P}{m_{3/2}} ,\\
\label{threshmssq1}
\Delta m_i^2&=&M_0^2\sum_a 2 c_a N
\frac{g_a^4 (M_{\rm mess}) }{(16\pi^2)^2} \left [\alpha_m  (1+\alpha_g)  \ln \frac{M_P}{m_{3/2}}\right ]^2.
\end{eqnarray}
The parameters of the model are the mass scales $M_0$ and $M_{\rm mess}$, as well as the dimensionless quantities $\alpha_m$, $\alpha_g$, the number of $SU(5)$ messenger pairs $N$, the modular weights $n_i$,  $\tan\beta$, and ${\rm sign}\mu$ (the model-dependent $\mu$ and $B\mu$ parameters are replaced as usual by the $Z$ boson mass, $\tan\beta$, and the sign of $\mu$).  Here, we will fix the modular weights to the standard values $n_i=1/2$ for all SM matter fields, and $n_i=1$ for the MSSM Higgs doublets.  With this choice, we see that in deflected mirage mediation, in addition to the usual $\tan\beta$ and ${\rm sign}\mu$, there are the continuous parameters $M_0$ and $M_{\rm mess}$, the discrete parameter $N$, and the parameters $\alpha_m$ and $\alpha_g$, which take on discrete values within a particular string framework but can be taken to be continuous parameters within a purely phenomenological approach.  

Hence, deflected mirage mediation models in this class have 6 parameters plus one sign.  This is to be compared with the 4 continous parameters (the universal scalar mass $M_0$, the universal gaugino mass $M_{1/2}$, and the universal trilinear scalar coupling $A_0$, and $\tan\beta$) plus one sign (the sign of $\mu$) in mSUGRA/CMSSM models.  We will see that despite its relatively small extension of parameters compared to mSUGRA/CMSSM models, deflected mirage mediation provides a rich framework for LHC phenomenology.

%---------------------
\section{(Deflected) Mirage Unification and Particle Mass Spectra}
\label{sec:scan}

\begin{figure}[!h]
\begin{center}
\subfigure[]{\includegraphics[scale=0.35,angle=0]{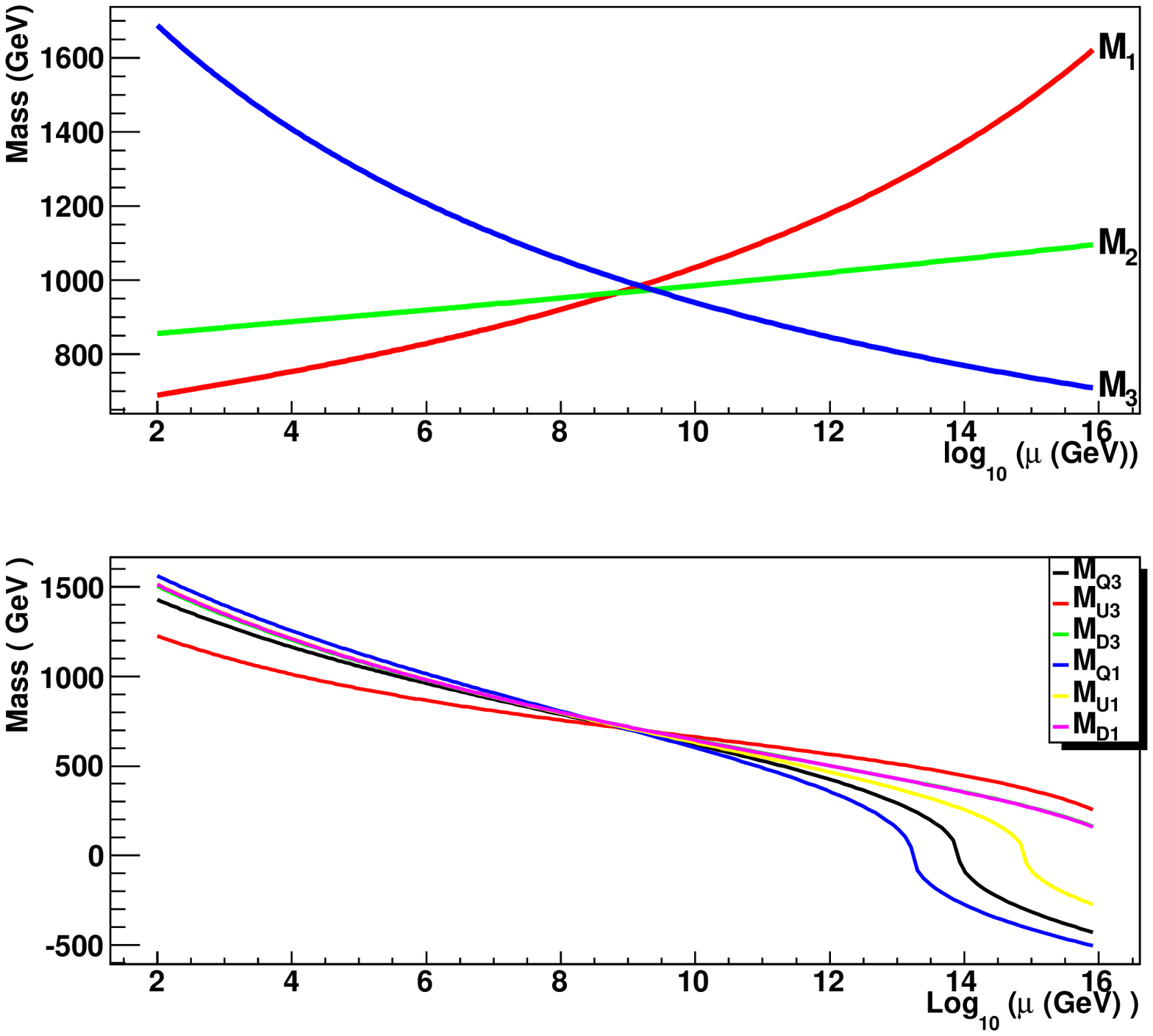}}
\subfigure[]{\includegraphics[scale=0.4,angle=0]{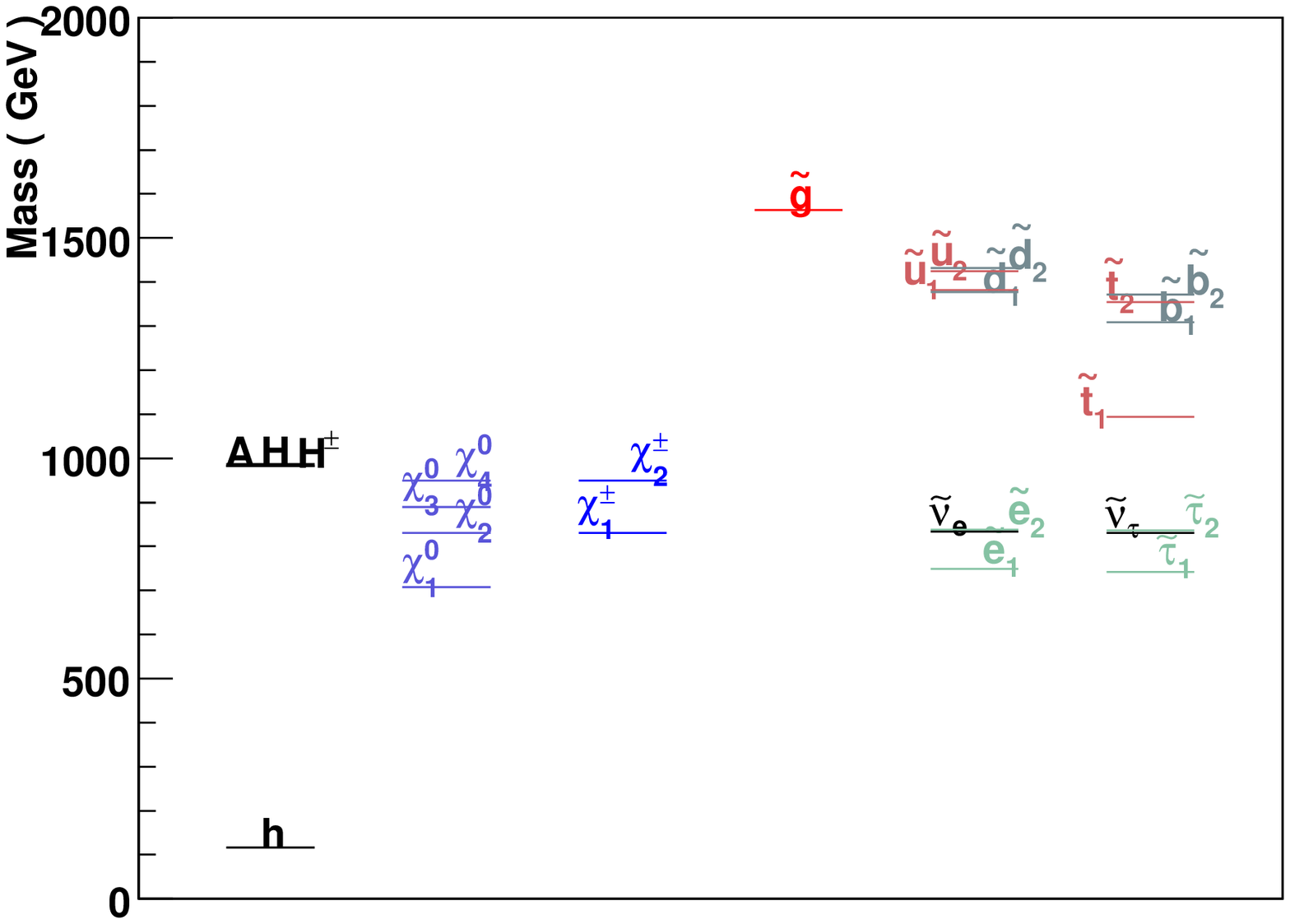}}
\caption{\textbf{Mirage unification}. The renormalization group evolution of the gauginos and soft mass-squared parameters (a) and the particle mass spectrum at low energies (b) for a pure mirage mediation scenario with $\alpha_m=1$, $M_0= 1$ TeV, $\tan\beta=10$, and $\mu>0$.}
\end{center}
\label{miragefig1}
\end{figure}

To compare deflected mirage mediation to mirage mediation, we provide here a discussion of the prototypical feature of mirage mediation, which is the phenomenon of mirage unification and its resulting profound impact on the low energy spectrum (see also \cite{Everett:2008qy,Everett:2008ey,Choi:2009jn} for previous discussions).  Mirage mediation is thus named because at the one-loop order, the soft terms unify not at the unification scale $M_G\sim 10^{16}$ GeV as in the case of mSUGRA/CMSSM models, but rather at a ``mirage" scale \cite{Choi:2004sx,Choi:2005ge,Choi:2005uz}:
\begin{equation}
\label{mirage0}
M_{\rm mir}=M_G \left (\frac{m_{3/2}}{M_{P}}\right )^{\alpha_m/2},
\end{equation}
where $\alpha_m$ is the ratio of anomaly mediation to gravity mediation terms given in Eq.~(\ref{param1}).  For the KKLT value of $\alpha_m=1$, the mirage unification occurs at $\sim10^{10}$ GeV, as shown in Fig.~\ref{miragefig1}.
Here we have included two-loop effects in the running, which spoil the precise unification, though the general features are maintained.  The mass spectrum, also given in Fig.~\ref{miragefig1}, shows that this model has a relatively heavy gluino, but has a slightly compressed spectrum with respect to corresponding mSUGRA/CMSSM models.   Smaller values of $\alpha_m$ have mirage unification scales closer to $M_G$ (with $M_{\rm mir}=M_G$ in the limit of $\alpha_m=0$), while larger values of $\alpha_m$ result in lower mirage unification scales. The case of $\alpha_m=2$ results in a mirage unification scale at TeV energies, resulting in a highly compressed superpartner spectrum and a reduced little hierarchy problem  \cite{miragepheno,littlehierarchy}.

In deflected mirage mediation, mirage unification is maintained in the gaugino sector (though lost in general in the scalar sector), as follows \cite{Everett:2008qy,Everett:2008ey}:
\begin{equation}
\label{miraged}
M_{\rm mir}=M_G \left (\frac{m_{3/2}}{M_{Pl}}\right )^{\alpha_m \rho/2},
\end{equation}
with the parameter $\rho$ given by
\begin{equation}
\label{rhodef}
\rho= \left (1+ \frac{2Ng_0^2}{16\pi^2}  \ln \frac{M_{\rm GUT} }{M_{\rm mess}}\right )
\left (1- \frac{\alpha_{\rm m} \alpha_{\rm g} Ng_0^2}{16\pi^2} 
 \ln \frac{M_P}{m_{3/2}}\right )^{-1}.
\end{equation}
Eqs.~(\ref{miraged})--({\ref{rhodef}) show that the mirage mediation limit is obtained for $N=0$ ({\it i.e.}, the absence of messengers).  If $\alpha_g=0$, corresponding to vanishing gauge mediation contributions, the messengers still contribute to anomaly mediation, as reflected in Eq.~(\ref{rhodef}).

The gaugino mirage unification scale in deflected mirage mediation can change widely from the pure mirage mediation case with fixed $\alpha_m$, with the details depending on the values of $\alpha_g$, $N$,  and the messenger scale $M_{\rm mess}$.   For nonzero $N$, the presence of the messengers results in $\rho > 1$ when $\alpha_g=0$, such that the mirage unification scale is lowered compared to that of the pure mirage mediation case.  For fixed $\alpha_m$, $N$, and $M_{\rm mess}$, the effects of a nonzero $\alpha_g$ are straightforward to understand: for $\alpha_g>0$, $\rho$ increases and $M_{\rm mir}$ is lowered, while for $\alpha_g<0$, $\rho$ decreases and $M_{\rm mir}$ is correspondingly increased.  

As an example of a model with gaugino mass mirage unification near the TeV scale, in Fig.~\ref{dmiragefig} we show  the renormalization group running of the gaugino and soft scalar mass-squares, as well as the particle mass spectrum at the TeV scale, for the case of $\alpha_g=1$ and $M_{\rm mess}=10^{12}$ GeV.
\begin{figure}
%[!h]
\begin{center}
\subfigure[]{\includegraphics[scale=0.35,angle=0]{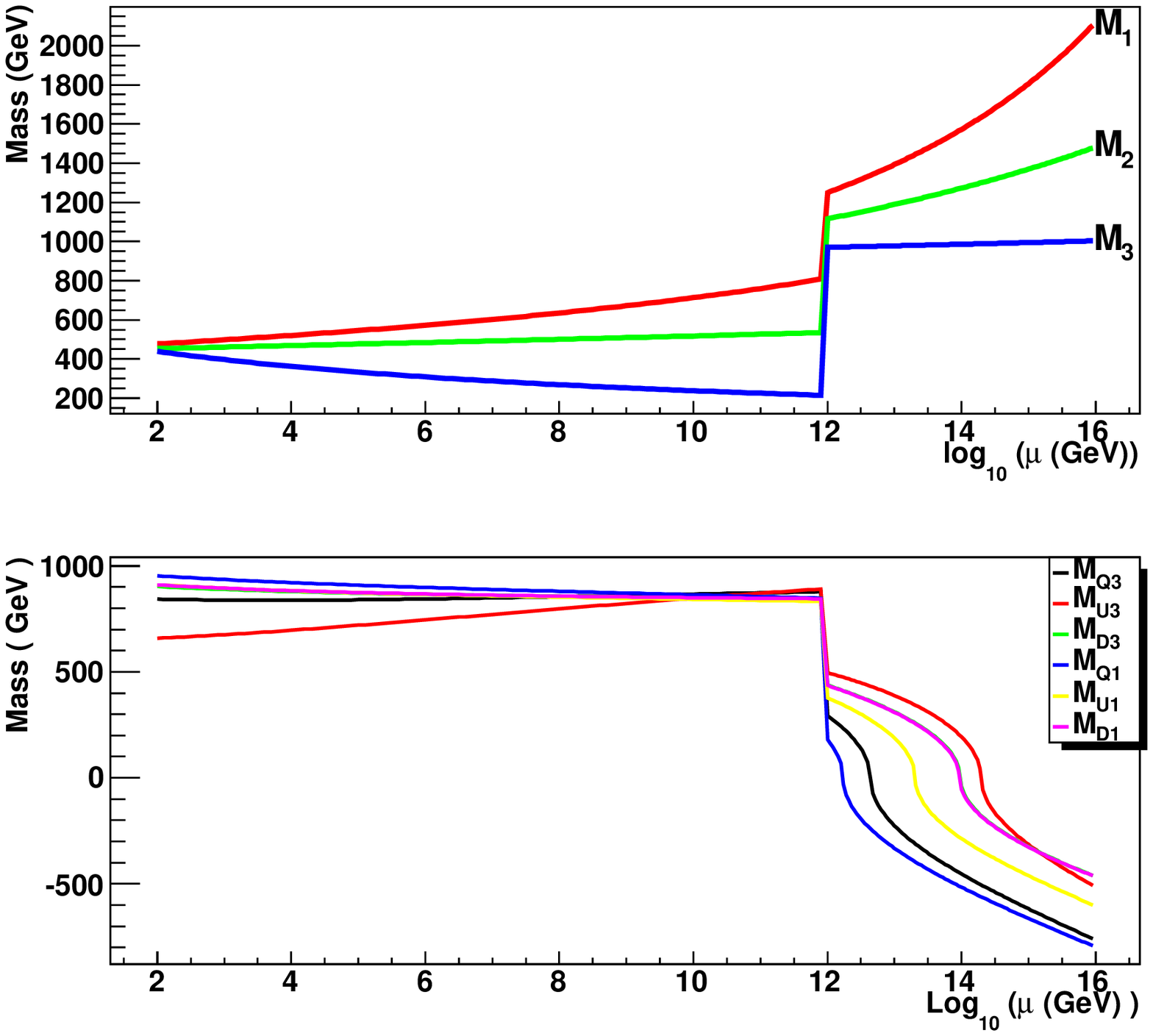}}
\subfigure[]{\includegraphics[scale=0.4,angle=0]{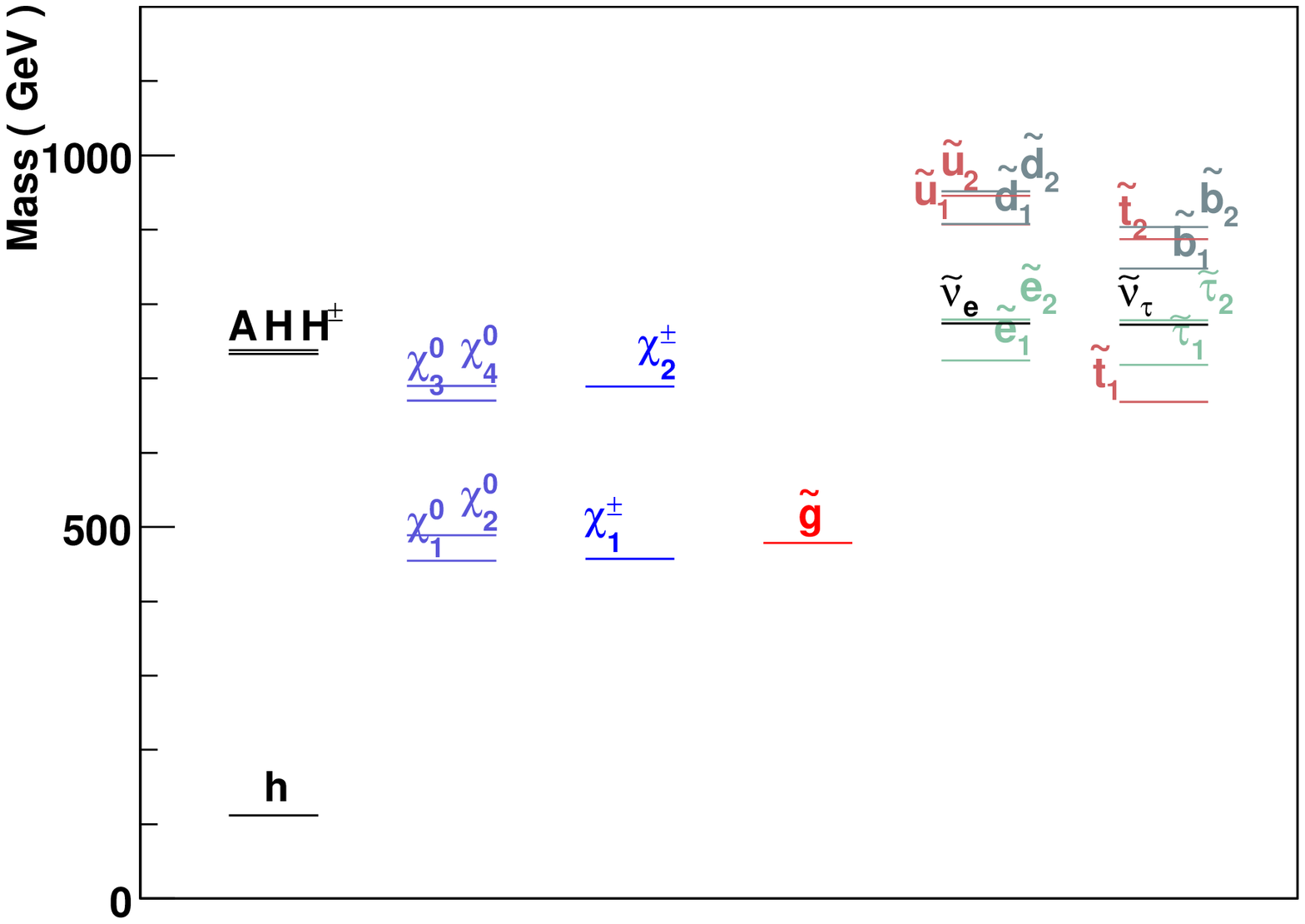}}
\caption{\textbf{Deflected mirage unification}. The renormalization group evolution of the gauginos and soft mass-squared parameters (a) and the mass spectrum (b) for a deflected mirage mediation model with $\alpha_m=1$, $\alpha_g=1$, $M_{\rm mess}=10^{12}$ GeV, $N=3$, $M_0= 1$ TeV, $\tan\beta=10$, and $\mu>0$.}
\end{center}
\label{dmiragefig}
\end{figure}
Clearly, the gaugino mass spectrum is highly squeezed, with a very light gluino (the lightest of the colored superpartners). Indeed, for $N = 3$ the beta-function coefficient for the soft supersymmetry breaking parameter $M_3$ vanishes at one loop above the messenger scale, and hence the soft mass for the gluino then runs very little between the high energy input scale and
the messenger scale.  Therefore, the gluino mass is much smaller for
these cases than for analogous pure mirage mediation models.  Furthermore, in contrast to the pure mirage mediation case with $\alpha_m\simeq 2$, which also has mirage unification near the TeV scale, the deflected mirage mediation scenarios with low-scale mirage unification have heavier scalars.  The full superpartner spectrum is thus stretched with respect to the pure mediation scenario due to the effects of gauge mediation.

In contrast, we show in Fig.~\ref{smallthresholds}
a deflected mirage mediation scenario with the same parameters as that of the previous figure, except that $\alpha_g=-0.5$.  In this case, there is a much higher gaugino mirage unification scale that is close to that of the pure mirage mediation limit, since with this choice of parameters $\rho \approx 1$. As a result, the particle mass spectrum strongly resembles that of Fig.~1, with heavier colored superpartners and no strong degeneracy between the lightest chargino and neutralino.   This feature depends on the messenger scale; for $\alpha_g<0$, there are always pairs of $\alpha_g$ and $M_{\rm mess}$ for which $\rho=1$ and thus $M_{\rm mir}$ is given by the pure mirage mediation limit.
 \begin{figure}
 %[!h]
 \label{smallthresholds}
\begin{center}
\subfigure[]{\includegraphics[scale=0.35,angle=0]{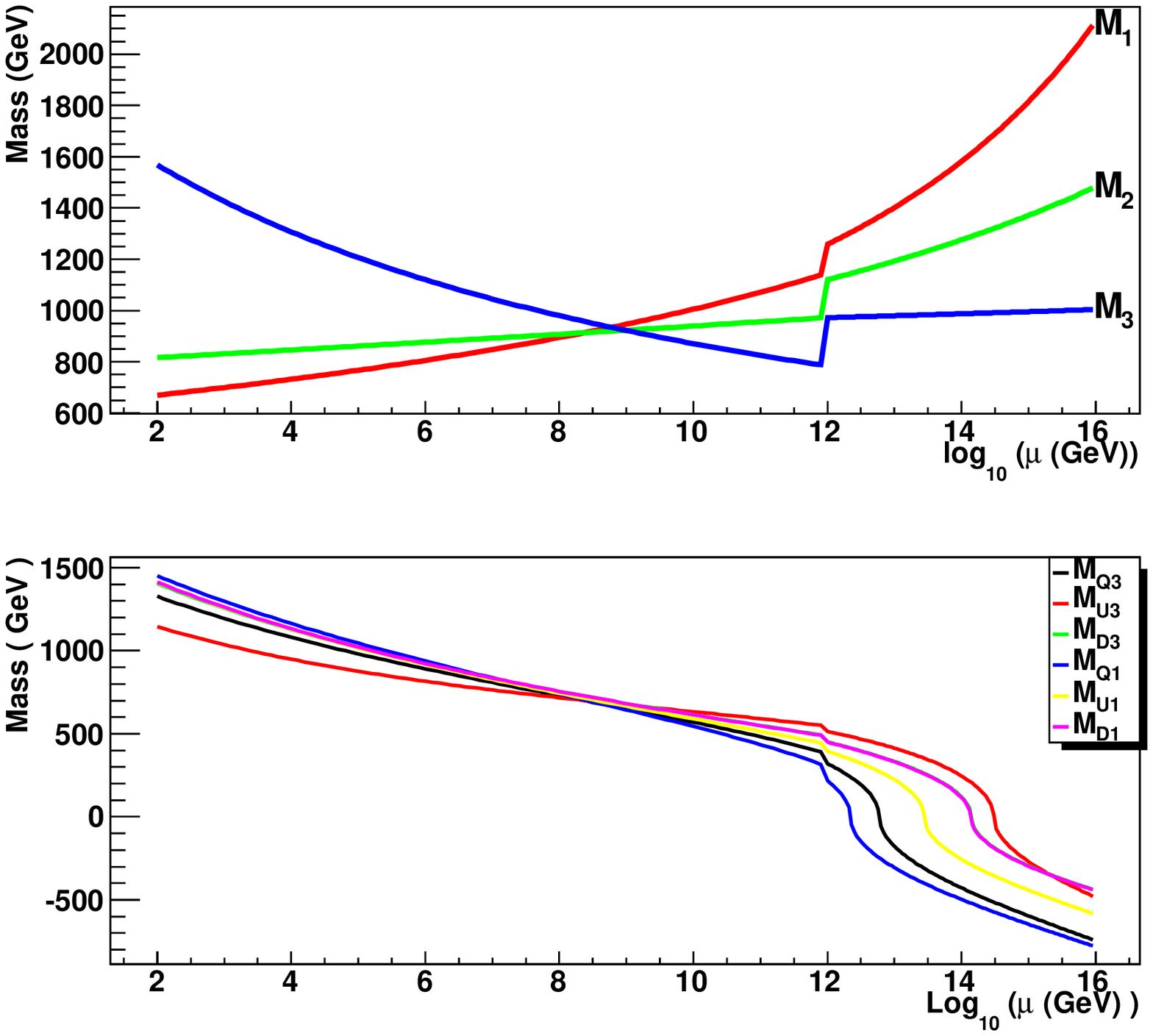}}
\subfigure[]{\includegraphics[scale=0.4,angle=0]{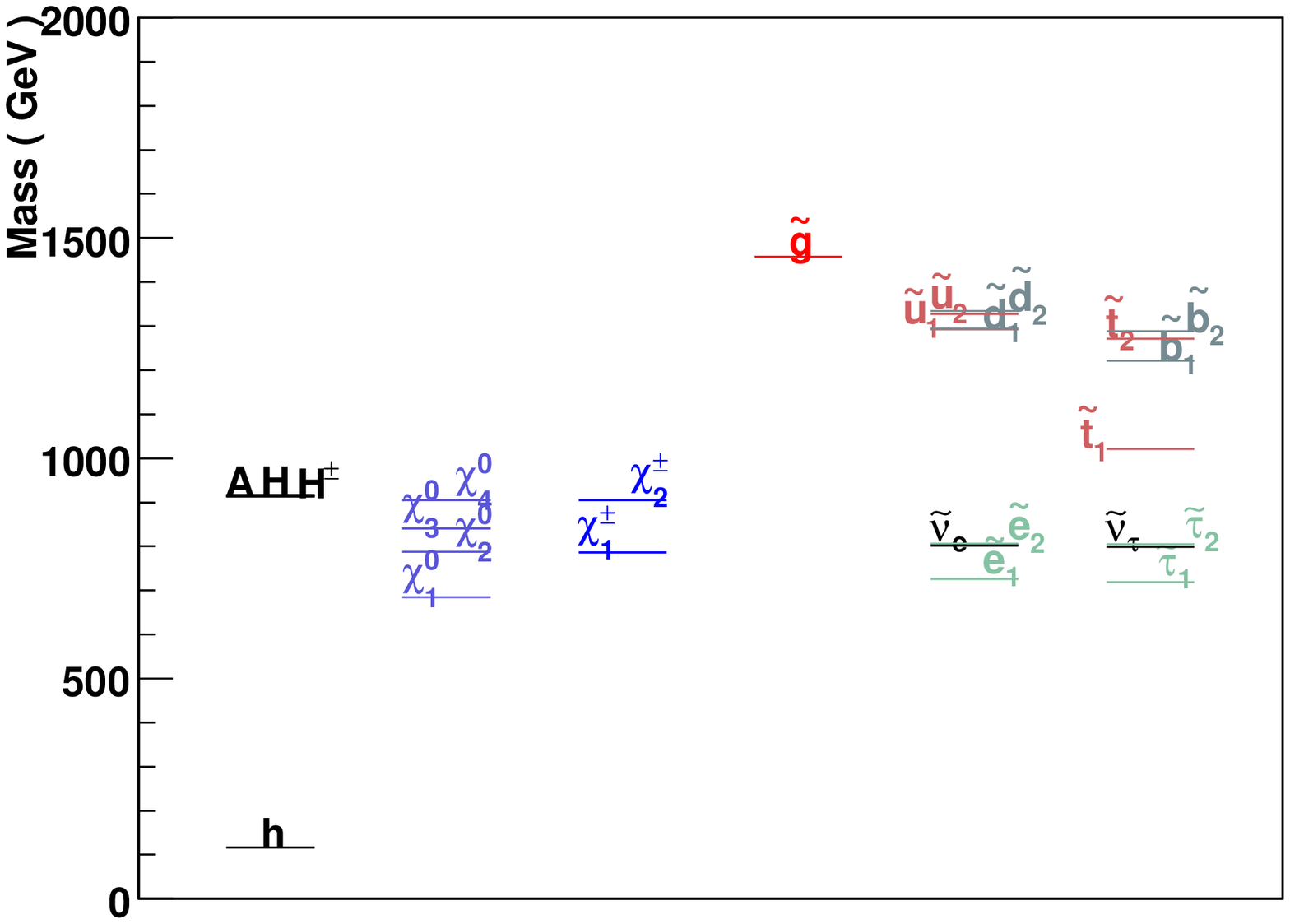}}
\caption{\textbf{Deflected mirage unification}. The renormalization group evolution of the gauginos and soft mass-squared parameters (a) and the mass spectrum (b) for a deflected mirage mediation model with $\alpha_m=1$, $\alpha_g=-0.5$, $M_{\rm mess}=10^{12}$ GeV, $N=3$, $M_0= 1$ TeV, $\tan\beta=10$, and $\mu>0$.}
\end{center}
\end{figure}
 \begin{figure}
 %[!h]
 \label{winoLSPcase}
\begin{center}
\subfigure[]{\includegraphics[scale=0.35,angle=0]{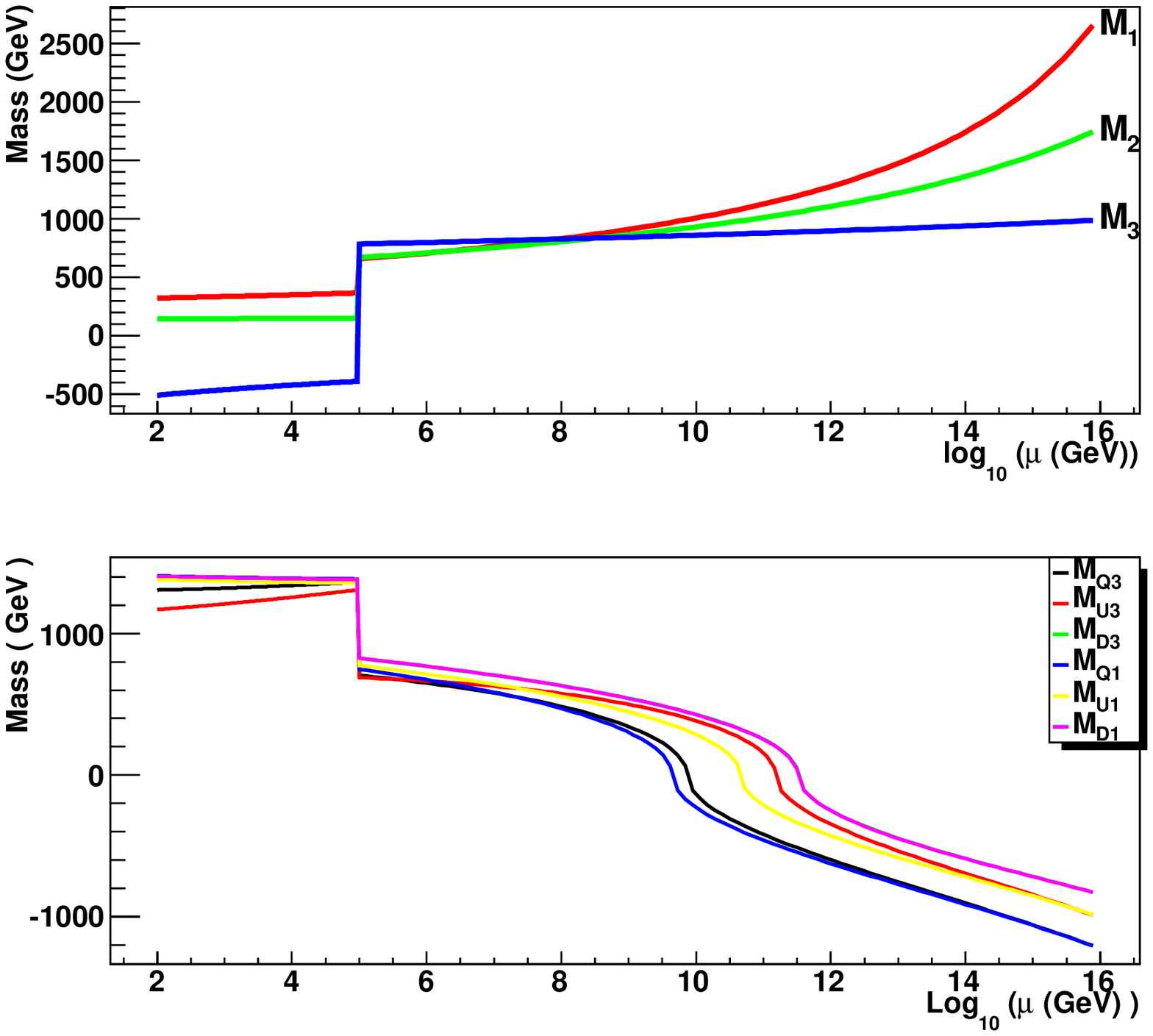}}
\subfigure[]{\includegraphics[scale=0.4,angle=0]{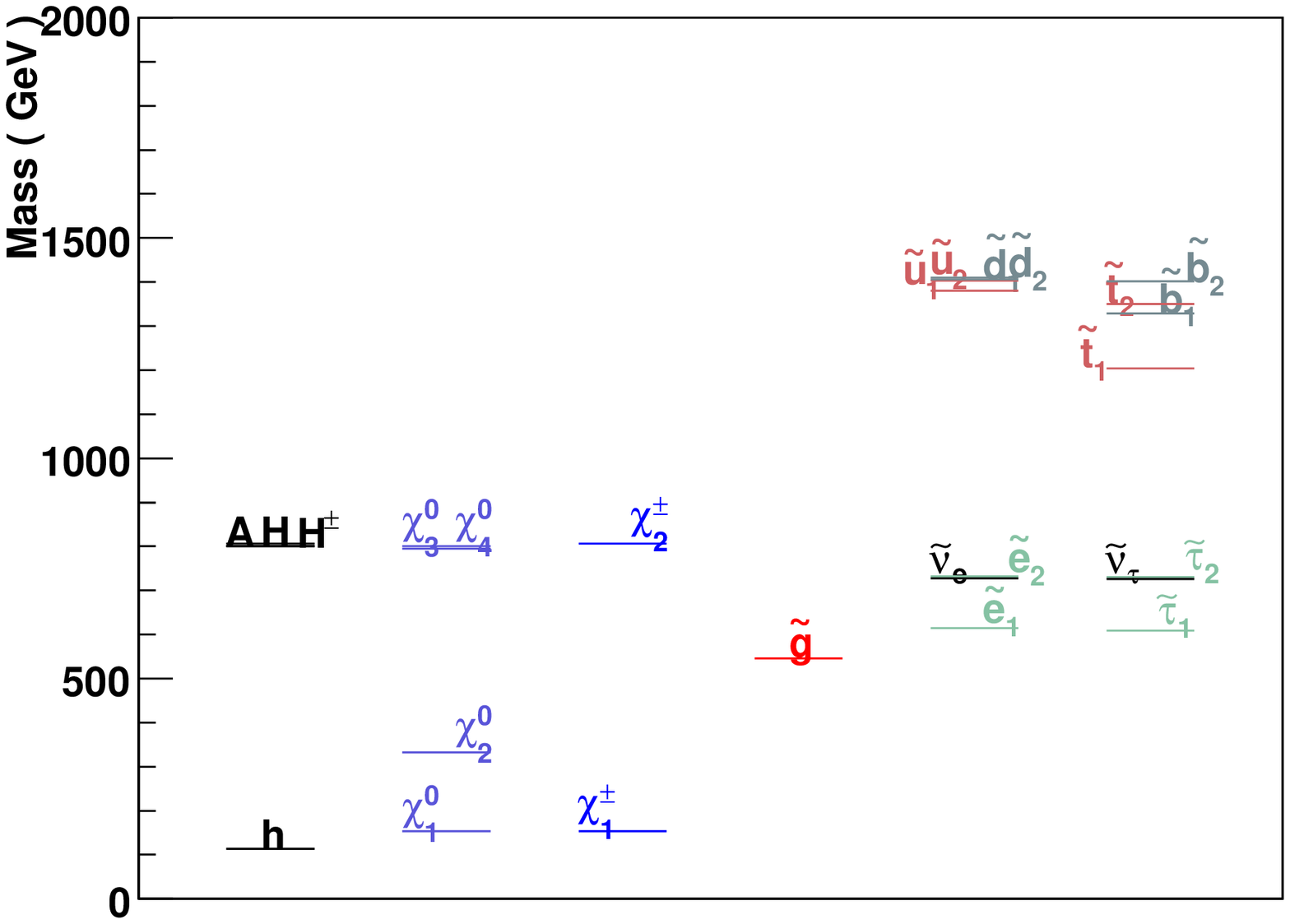}}
\caption{\textbf{Deflected mirage unification}. The renormalization group evolution of the gauginos and soft mass-squared parameters (a) and the mass spectrum  (b) for a deflected mirage mediation scenario with $\alpha_m=1$, $\alpha_g=1$, $M_{\rm mess}=10^{5}$ GeV, $N=3$, $M_0= 1$ TeV, $\tan\beta=10$, and $\mu>0$.}
\end{center}
\end{figure}

Furthermore, if the messenger scale is below the mirage unification scale as given in the pure mirage mediation case by Eq.~(\ref{mirage0}), the mirage unification behavior is maintained not only for the gauginos, but also for the scalars (though at a different scale than the pure mirage mediation limit due to the presence of the messengers).  For $\alpha_g>0$, we can find model points in which the gaugino mass unification scale is sub-TeV, resulting in a flipped gaugino mass spectrum in which the wino is the lightest superpartner, similar to the gaugino pattern in the pure anomaly mediation limit. An example is shown in 
Fig.~\ref{winoLSPcase}, in which all parameters are the same as in Fig.~2, except that 
$M_{\rm mess}=10^{5}$ GeV.

We see, therefore, that deflected mirage mediation models roughly can be categorized according to their values of $\alpha_g$ as follows: the case of large threshold effects ($\alpha_g>0$, nonperturbative stabilization), and the case of small threshold effects ($\alpha_g<0$, radiative or higher-dimensional stabilization).  This can be easily understood from Eqs.~(\ref{gaugino1})--(\ref{threshmssq1}); for $\alpha_g>0$, $F^X/X$ and $F^C/C$ have the same sign, and thus the threshold effects for gauge and anomaly mediation constructively interfere, while for $\alpha_g<0$, the threshold effects destructively interfere.   For small thresholds, the particle mass spectra are typically similar to a corresponding pure mirage mediation spectrum, while for large thresholds, deflected mirage mediation can result in significantly different mass spectra with non-standard gaugino mass patterns characterized by a squeezed gaugino mass spectrum with relatively light gluinos and a lightest superpartner with a mixed wino-bino-Higgsino content, similar to gauge messenger models \cite{Dermisek:2006qj,Bae:2007pa}.  This is phenomenologically interesting for a variety of reasons, including fine-tuning
considerations~\cite{littlehierarchy},
LHC
signatures~\cite{Baer:2006id,Choi:2007ka,Cho:2007fg,Baer:2007eh},
and dark matter
signals~\cite{BirkedalHansen:2001is,BirkedalHansen:2002am,Choi:2006im,ArkaniHamed:2006mb,King:2006tf,Holmes:2009mx}.
However, the thresholds must not be too large, otherwise the gluino can become the lightest superpartner.  Therefore, viable deflected mirage mediation models with large threshold effects have bounds on the allowed range of $\alpha_g$, depending on the other parameters.     The examples we have shown here, while not chosen to optimize the dark matter predictions, encompass all of these possibilities. 

%---------------------
\section{Collider Phenomenology}
\label{sec:collider}

The soft term expressions in Eqs.~(\ref{gaugino1})--(\ref{threshmssq1}) have
sufficient complexity to produce a wide variety of possible
low-energy superpartner spectra. The analysis of
Section~\ref{sec:scan} gives evidence of the wide diversity of
outcomes which can arise. In this section we will turn our attention
to how the addition of gauge mediation to the mirage pattern of
particle masses influences the collider phenomenology of this class
of models.
%%%%%%%%%%%%%%%%%%%%%%%%%%%%%%%%

\subsection{Comparison of Mirage and Deflected Mirage Mediation}
\label{benchmarks}

We will begin by considering a set of benchmark points designed to
illustrate some of the differences between pure mirage
mediation models and deflected mirage mediation models, focusing on the case of large thresholds ($\alpha_g >0$) as described in Section~\ref{sec:scan}.  The benchmark models were chosen such that the deflected mirage mediation models satisfy the Higgs mass bound and the dark matter relic density constraints; note that the pure mirage mediation models given here do not satisfy these constraints, but are shown for sake of comparison.  
The high scale
input parameters which define these points are collected in
Table~\ref{tbl:benchmarks1}, along with the physical particle masses at the TeV scale and the composition of the lightest neutralino, which is the lightest superpartner (LSP) in these models. The six model points are grouped into three
pairs, with the first model in each pair being a pure mirage mediation 
model with the number of vector-like $\mathbf{5} + \bar{\mathbf{5}}$
messenger fields set to $N =0$, while the second model is a
deflected mirage point for which we have taken $N = 3$. For all of
the model points in Table~\ref{tbl:benchmarks1}, we have set $\tan\beta =
10$ and have chosen modular weights $n_i = 1/2$ for the matter fields and $n_i = 1$ for the Higgs fields, as described previously.

%========= table of benchmark properties ==========
\begin{table}[t]
{\begin{center}
\begin{tabular}{|l||c|c||c|c||c|c|}\cline{1-7}
% POINT 1 and 2 = TeV SCALE UNIFICATION
% POINT 3 and 4 =
% POINT 5 and 6 =
%
 & Point 1 & Point 2 & Point 3 & Point 4 & Point 5 & Point 6 \\
\cline{1-7}
$\alpha_m$ & 1.9 & 1.0 & 0.6 & 0.6 & 1.0 & 1.0 \\
$\alpha_g$ & 0 & 0.5 & 0 & 1.0 & 0 & 0.2 \\
$M_0$ & 1000 %GeV 
& 1000 & 1000 & 1000 & 800 & 800 \\
$M_{\rm mess}$ & NA & $10^{10}$  & NA & $10^{7}$  & NA & $10^{10}$  \\
$N$ & 0 & 3 & 0 & 3 & 0 & 3 \\
\hline
%\cline{1-7}
\hline
$m_{\wtd{N}_{1}}$ & 236 & 493 & 602 & 322 & 562 & 424\\
$m_{\wtd{N}_{2}}$ & 247 & 516 & 848 & 329 & 660 & 452\\
$m_{\wtd{N}_{3}}$ & 936 & 698 & 1114 & 943 & 725 & 569\\
$m_{\wtd{N}_{4}}$ & 954 & 718 & 1127 & 946 & 779 & 601\\
$m_{\wtd{C}_{1}^{\pm}}$ & 243 & 498 & 848 & 328 & 658 & 441\\
$m_{\wtd{C}_{2}^{\pm}}$ & 937 & 718 & 1133 & 952 & 779 & 599\\
\cline{1-7}
$m_{\tilde{\tau}_{1}}$ & 676 & 700 & 763 & 717 & 594 & 556\\
$m_{\tilde{\tau}_{2}}$ & 687 & 760 & 892 & 808 & 672 & 605\\
$m_{\tilde{\mu}_{R}}$, $m_{\tilde{e}_{R}}$ & 679 & 706 & 773 & 726 & 600 & 562\\
$m_{\tilde{\mu}_{L}}$, $m_{\tilde{e}_{L}}$ & 685 & 761 & 894 & 810 & 672 & 605\\
\cline{1-7}
$m_{\tilde{t}_{1}}$ & 620 & 687 & 1278 & 803 & 875 & 560\\
$m_{\tilde{t}_{2}}$ & 829 & 913 & 1579 & 1091 & 1115 & 777\\
$m_{\tilde{b}_{1}}$ & 716 & 865 & 1542 & 1055 & 1062 & 713\\
$m_{\tilde{b}_{2}}$ & 751 & 936 & 1624 & 1153 & 1115 & 773\\
$m_{\tilde{c}_{R}}$, $m_{\tilde{u}_{R}}$ & 733 & 933 & 1639 & 1160 & 1121 & 769\\
$m_{\tilde{c}_{L}}$, $m_{\tilde{u}_{L}}$ & 713 & 962 & 1695 & 1204 & 1155 & 788\\
$m_{\tilde{s}_{R}}$, $m_{\tilde{d}_{R}}$ & 751 & 940 & 1633 & 1162 & 1119 & 777\\
$m_{\tilde{s}_{L}}$, $m_{\tilde{d}_{L}}$ & 721 & 968 & 1702 & 1210 & 1162 & 794\\
$m_{\tilde{g}}$ & 979 & 603 & 1816 & 431 & 1266 & 646\\ \hline \hline
LSP Bino \% & 0.2\% & 19.1\% & 99.5\% & 82.0\% & 93.1\% & 52.5\% \\
LSP Wino \% & 0.8\% & 70.0\% & 0.0\% & 17.0\% & 0.9\% & 30.1\% \\
LSP Higgsino \% & 99.0\% & 10.9\% & 0.5\% & 1.0\% & 6.0\% & 17.4\% \\
\cline{1-7}
\end{tabular}
\end{center}}
{\caption{\label{tbl:benchmarks1}\footnotesize{\bf Input Parameters, Physical Masses, and LSP Composition for Benchmark Models}. 
The first model in each pair is a mirage mediation model and the second is a deflected mirage mediation model with $N = 3$.  All masses are given in GeV.
Low energy physical masses are given at the scale~1~TeV. 
}}
\end{table}
%=================================================================

The first pair are models for which there is mirage unification at the TeV scale in the gaugino sector.  Point~1 is a pure mirage mediation model with $\alpha_m \simeq 2$, and hence the scalars are also unified at TeV energies in this case.    Point~2 is a deflected mirage mediation model with $\alpha_m=1$ and $\alpha_g>0$,  which has TeV-scale gaugino mirage unification.  The unification is not exact due to two-loop effects: for Point~2 the three gaugino soft masses at the electroweak scale are $M_1 = 500$ GeV, $M_2= 494$ GeV, and $M_3= 574$ GeV, while for Point~1 they are 
$M_1 = 929$ GeV, $M_2= 929$ GeV, and $M_3= 1062$ GeV.  For Point~1, the electroweak symmetry breaking conditions constrain the $\mu$-parameter to a relatively small value of $\mu =
239\GeV$. The approximate unification at the electroweak scale is then between the gluino and the {\em heavier} pair of
neutralinos and heavier chargino. The lighter pair of neutralinos
are mostly Higgsino-like, as indicated by the LSP composition given
in Table~\ref{tbl:benchmarks1}. For Point~2, however, the gaugino spectrum is
compressed and the gluino approximately unifies with the entire
ensemble of neutral and charged gauginos.

The last two pairs show the effects of keeping $\alpha_m$ fixed and adding nonvanishing gauge mediation contributions by including messenger fields and a nonvanishing $\alpha_g$.
The pair represented by Points~3 and~4 is designed to show the impact of adding the effects of gauge
mediation on the resulting gaugino masses. Both points have $M_0 =
1\TeV$ and $\alpha_m = 0.6$, yet the two cases have very different
phenomenology. The spectrum for the pure mirage mediation point is
similar to typical mSUGRA/CMSSM models, with a
bino-like LSP, large mass gap between the lightest and second
lightest neutralinos, gluinos and squarks of roughly comparable
size, and a relatively light set of sleptons.
In contrast, the deflected mirage mediation point has a mixed
bino/wino-like LSP, a degenerate trio of $\wtd{N}_1$, $\wtd{N}_2$
and $\wtd{C}_1$, and a very light gluino
relative to the squarks and sleptons. 
The final pair, represented by Points~5 and~6, are models which again have the same values of $\alpha_m=1$ and $M_0=800$ GeV.  These models have very similar spectra for the light superpartners, but very different values for the gluino and squark masses and the $\mu$-parameter.  The pure mirage mediation point (Point 5) has a predominantly bino-like LSP, the deflected mirage mediation point (Point 6) has a neutralino LSP which is a mixed bino-wino-Higgsino state.

To study the collider signatures of these points at the LHC, 50,000~events were generated for each model at $\sqrt{s}=14$~TeV using {\tt PYTHIA
6.4}~\cite{Sjostrand:2006za}. Generated events are passed to {\tt
PGS4}~\cite{PGS} to simulate the detector response. Events are
analyzed using the {\tt PGS4} level one triggers, designed to mimic
the CMS trigger tables~\cite{Ball:2007zza}. Object-level
post-trigger cuts were also imposed. We require all photons,
electrons, muons and taus to have transverse momentum $p_T\geq 10$
GeV and $|\eta|<2.4$ and we require hadronic jets to satisfy
$|\eta|<3$. Additional post-trigger level cuts were implemented for
specific analyses, as described below.

%========= LHC discovery modes, benchmarks ==========
\begin{table}[t]
{\begin{center}
\begin{tabular}{|l||c|c||c|c||c|c|}\cline{1-7}
 & Point 1 & Point 2 & Point 3 & Point 4 & Point 5 & Point 6 \\
\cline{1-7}
$\sigma_{\SUSY}$ (pb) & 5.86 & 10.86 & 0.045 & 44.71 & 0.58 & 13.26 \\
Trigger Efficiency & 84.8\% & 78.9\% & 99.3\% & 59.4\% & 98.4\% & 87.1\% \\
\cline{1-7}
%
%\multicolumn{1}{c}{ } &
\multicolumn{7}{c}{Counts per 50,000 Events} \\
\hline
Multijet & 5064 & 1250 & 10113 & 579 & 4645 & 1246 \\
1~Lepton & 694 & 69 & 3861 & 19 & 4266 & 445 \\
OS~Dilepton & 28 & 0 & 370 & 0 & 1623 & 9 \\
SS~Dilepton & 3 & 0 & 124 & 0 & 201 & 3 \\
Trilepton & 0 & 0 & 70 & 0 & 388 & 1 \\
\cline{1-7}
\end{tabular}
\end{center}}
{\caption{\label{tbl:xsecbench}\footnotesize{\bf Gross LHC Features
for Benchmark Points}. The trigger efficiency is here computed using
the level one trigger table of {\tt PGS4}. The number of events
passing our selection criteria in the multijet, single lepton plus
jets, opposite-sign dilepton plus jets, same-sign dilepton plus
jets, and trilepton plus jets channels are given for 50,000
generated events.}}
\end{table}
%========= LHC discovery modes, benchmarks ==========
\begin{table}[t]
{\begin{center}
\begin{tabular}{|l||c|c||c|c||c|c|}\cline{1-7}
 & Point 1 & Point 2 & Point 3 & Point 4 & Point 5 & Point 6 \\
\cline{1-7}
Multijet & 0.17 & 0.37 & 70.30 & 0.20 & 1.90 & 0.30 \\
1~Lepton & 2.18 & 64.20 & -- & 49.90 & 5.87 & 1.04 \\
OS~Dilepton & 68.80 & -- & -- & -- & 2.08 & -- \\
SS~Dilepton & -- & -- & -- & -- & 11.91 & -- \\
Trilepton & -- & -- & -- & -- & 2.94 & -- \\
\cline{1-7}
\end{tabular}
\end{center}}
{\caption{\label{tbl:sigbench}\footnotesize{\bf Necessary Integrated
Luminosity for $5\sigma$ Discovery in Selected Channels}. The
integrated luminosity (in fb$^{-1}$) at $\sqrt{s}=14\TeV$ to produce
a $5\sigma$ excess over Standard Model backgrounds is given for all
cases in which $\mathcal{L}_{\rm int} \leq 100\, {\rm fb}^{-1}$. We
require a minimum of 100 signal events in the no-lepton and single
lepton channels, and a minimum of ten signal events in the
multi-lepton channels.}}
\end{table}
%=================================================================
%=================================================================

The total cross section for superpartner production  is given in
 Table~\ref{tbl:xsecbench} for each 
benchmark model. To a first approximation the total cross section
is dependent solely on the size of the gluino mass and thus
deflected mirage mediation models offer the prospect of larger LHC signals relative to comparable pure mirage mediation models. The trigger
efficiency is estimated using the level one trigger table of {\tt
PGS4} and represents the fraction of the 50,000 generated events
that passed the trigger criteria. As we will see below, however, the
actual number of events that pass post-trigger cuts will often be 
much smaller.

The addition of the gauge messenger sector has its most striking
effect in the pair of Points~3 and~4, producing a difference in the
total production cross section of almost three orders of magnitude.
For Point~3, collecting 50,000 signal events will require
over 1000 fb$^{-1}$ of integrated luminosity, while Point~4 achieves this in just over 1 fb$^{-1}$. For smaller values of $N$, 
the gluino mass would be
larger and hence the expected signal size would diminish.
Triggering efficiencies are generally slightly better for models with a less compressed gaugino mass spectrum.   This results in slightly harder leptonic decay products at the final
stages of cascade decays of gluinos and squarks. The {\tt PGS4}
default level one trigger criteria requires leptons ($e^{\pm}$ and
$\mu^{\pm}$) to have $p_T \geq 10\GeV$ in the dilepton channel, $p_T
\geq 15\GeV$ for an isolated lepton produced with a tau, and $p_T
\geq 20\GeV$ for a single isolated lepton produced in association
with hard jets. In addition to this trigger requirement, standard
supersymmetry search algorithms involving jets, leptons and missing
transverse energy generally also demand minimum $p_T$ values for
leptonic objects.

To demonstrate the differences between deflected mirage mediation
models and their pure mirage mediation model analogs, we will here
concern ourselves  with counting observables associated
with traditional discovery channels for
supersymmetry~\cite{Branson:2001ak}, reserving a more detailed
analysis of collider signatures for the following subsection. These
five signatures are collected in Table~\ref{tbl:xsecbench} for
50,000 generated events at  $\sqrt{s} = 14\TeV$. These signatures
are defined as follows. All five require transverse sphericity $S_T
\geq 0.2$ and at least 250~GeV of $\not\!\!{E_T}$ except for the
trilepton signature, where only $\not\!\!{E_T} \geq 200 \,{\rm GeV}$
is required.
Multijets here refers to events with at least four jets with the
transverse momenta of the four leading jets satisfying $p_T \geq
\left(200,150,50,50\right)$ GeV, respectively. For this signature we
impose a veto on isolated leptons.
For the single lepton, opposite-sign dilepton, same-sign dilepton
and trilepton signatures we include only $e^\pm$ and $\mu^{\pm}$
final states and demand at least two jets with the leading jets
satisfying $p_T \geq \left(100,50\right)$ GeV, respectively.
The drastic reduction in the multijet and the leptonic signatures for the deflected
mirage mediation models seen in Table~\ref{tbl:xsecbench} is 
caused by the small mass gap between the LSP and either the gluino or the lightest chargino/second neutralino. This is also true of the TeV-scale mirage unification model of Point~1.

Since the total cross sections vary significantly between the benchmark models, a more relevant number for comparison of the discovery potential between model points is the amount of integrated luminosity
necessary to observe a clear excess of events over the Standard
Model background. For this we generated a sample of 5~fb$^{-1}$
Standard Model background events, consisting of Drell-Yan, QCD
dijet, $t\,\bar{t}$, $b\,\bar{b}$, $W$/$Z$+jets and diboson
production. Scaling the weight of this sample relative to the total
cross section in Table~\ref{tbl:xsecbench} we can determine when a
$S/\sqrt{B} = 5\sigma$ excess will be detectable at
$\sqrt{s}=14\TeV$. The results are given in
Table~\ref{tbl:sigbench}. Note that we only extrapolate the value of
$S/\sqrt{B}$ for cases where $\mathcal{L}_{\rm int} \leq 100\,{\rm
fb}^{-1}$ and we require at least 100~signal events for the multijet
and single-lepton channels and at least 10~signal events for the
multi-lepton channels. With the exception of Point~3, all of these
benchmark points will give clear signals in the multijet channel early in the high luminosity phase of the LHC with $\sqrt{s}=14\TeV$. Leptonic discovery
channels will generally take longer to observe. Point~5, despite its
modest production cross-section of 0.6~picobarns, gives sizable
signals in all leptonic channels within the first 10~fb$^{-1}$. This
is largely due to the mass ordering $m_{\wtd{N}_1} <
m_{\wtd{\tau}_1} < m_{\wtd{N}_2}$, which does not
appear in any of the deflected mirage mediation models considered here. While several points would produce $\order(1000)$ signal
events in 1~fb$^{-1}$ at $\sqrt{s}=7\TeV$, the reliance on multijet
+ $\not\!\!{E_T}$ channels and absence of strong leptonic signals
suggest that these points will be challenging to discover in the first year of LHC running.

\subsection{Influence of $\alpha_g$ on LHC Phenomenology}
\label{alphalines}

In this subsection, we wish to study in greater detail how the size
of the gauge-mediated contribution to soft supersymmetry breaking
affects the expected collider signatures at the LHC for deflected
mirage mediation scenarios.
%%%%%%%%%%%%%%%%%%%%%%%%%%%%%%%%
To do so we construct four model ``lines'' in which the various
parameters determining the soft supersymmetry breaking masses are
fixed at specific values, while allowing the parameter $\alpha_g$ to
vary. These points represent a variety of overall mass scales and
spectra. We summarize the relevant input parameters in
Table~\ref{tbl:lines}. For each case, we have chosen to fix
$N =3$,  $n_i = 1/2$ for the matter representations and $n_i = 1$ for the Higgs fields, and $\tan\beta=10$. Each
line involves five discrete points with $\alpha_g = \lbr
-1.0,\,-0.5,\,0.0,\,0.5,\,1.0\rbr$.

%========= table of model line parameters ==========
\begin{table}[t]
{\begin{center}
\begin{tabular}{|l||c|c|c||c|c|c|c|c|}
\multicolumn{1}{c}{ } & \multicolumn{3}{c}{Parameter Set} &
\multicolumn{5}{c}{$\alpha_g$
 Value} \\ \cline{1-9}
 & $\alpha_m$ & $M_0$ & $M_{\rm mess}$ & \parbox{1cm}{-1.0} & \parbox{1cm}{-0.5} &
 \parbox{1cm}{$\,\,\,0$} & \parbox{1cm}{0.5} & \parbox{1cm}{1.0} \\
\cline{1-9}
Line~A & $1$ & $2\TeV$ & $10^{12}\GeV$
& $\tilde{\tau}$~LSP & $\checkmark$ & $\checkmark$ & $\checkmark$ & $\checkmark\checkmark$ \\
Line~B & $1$ & $1\TeV$ & $10^{8}\GeV$
& $\checkmark$ & $\checkmark\checkmark$ & $\checkmark$ & $\tilde{g}$~LSP & $\tilde{g}$~LSP \\
Line~C & $0.771$ & $0.8\TeV$ & $10^{12}\GeV$
& $\checkmark$ & $\checkmark$ & $\checkmark$ & $\checkmark$ & $\checkmark$ \\
Line~D & $ 0.755$ & $0.4\TeV$ & $10^{12}\GeV$
& $\checkmark$ & $\checkmark$ & $\checkmark$ & $\checkmark$ & $\checkmark$ \\
\cline{1-9}
\end{tabular}
\end{center}}
{\caption{\label{tbl:lines}\footnotesize{\bf Input Parameters for
Benchmark Lines}. For each model line we begin with the input
parameter set indicated in the initial three columns. Five values of
the parameter $\alpha_g$ were studied, keeping other parameters
fixed. Points marked with a check-mark had acceptable low-energy
phenomenology. Points marked with the double check-mark were studied
in Ref.~\cite{Everett:2008qy}.}}
\end{table}%========= table of key masses ==========
\begin{table}%[t]
{\begin{center}
\begin{tabular}{|ll||c|c|c||c|c|c|c||c|c|c|}
\multicolumn{2}{c}{Model} &
\multicolumn{1}{c}{$m_{\tilde{g}}$} &
\multicolumn{1}{c}{$m_{\tilde{q}_1}$}&
\multicolumn{1}{c}{$m_{\tilde{t}_1}$}
& \multicolumn{1}{c}{$m_{\rm LSP}$} & \multicolumn{1}{c}{$\Delta^0$}
& \multicolumn{1}{c}{$\Delta^{\pm}$} &
\multicolumn{1}{c}{$m_{\tilde{\ell}_1}$}
 & \multicolumn{1}{c}{B\%} & \multicolumn{1}{c}{W\%}
& \multicolumn{1}{c}{H\%} \\
\cline{1-12}
\multicolumn{2}{l}{Line A} & \multicolumn{9}{c}{} \\
\cline{1-12}
 & A2 & 2828 & 2492 & 2027 & 1400 & 175 & 179 & 1445 & 96.4\% & 0.1\% & 3.5\% \\
 & A3 & 2260 & 2144 &1710 & 1265 & 132 & 132 & 1429 & 94.9\% & 0.4\% & 4.8\% \\
 & A4 & 1677 & 1895& 1479 & 1133 & 70 & 69 & 1427 & 94.1\% & 1.6\% & 4.3\% \\
 & A5 & 1045 & 1814& 1380 & 977 & 30 & 1.6 & 1441 & 3.6\% & 92.5\% & 3.9\% \\
\cline{1-12}
\multicolumn{2}{l}{Line B} & \multicolumn{9}{c}{} \\
\cline{1-12}
 & B1 & 1347 &1197&  942 & 663 & 84 & 80 & 686 & 88.7\% & 1.6\% & 9.6\% \\
 & B2 & 1038 & 1038& 785 & 595 & 54 & 49 & 679 & 85.7\% & 4.5\% & 9.7\% \\
 & B3 & 711 & 952& 707 & 525 & 20 & 11 & 677 & 51.4\% & 37.8\% & 13.4\% \\
\cline{1-12}
\multicolumn{2}{l}{Line C} & \multicolumn{9}{c}{} \\
\cline{1-12}
 & C1 & 1440 & 1277& 999 & 530 & 167 & 167 & 596 & 98.4\% & 0.1\% & 1.4\% \\
 & C2 & 1244 & 1133& 868 & 487 & 132 & 132 & 587 & 98.0\% & 0.2\% & 1.8\% \\
 & C3 & 1048 & 1003& 747 & 444 & 99 & 98 & 582 & 97.4\% & 0.4\% & 2.2\% \\
 & C4 & 847 & 894& 647 & 402 & 66 & 65 & 580 & 96.4\% & 1.0\% & 2.6\% \\
 & C5 & 640 & 818& 578 & 359 & 34 & 32 & 583 & 93.3\% & 3.7\% & 3.1\% \\
\cline{1-12}
\multicolumn{2}{l}{Line D} & \multicolumn{9}{c}{} \\
\cline{1-12}
 & D1 & 752 & 672& 496 & 254 & 75 & 73 & 297 & 94.3\% & 1.1\% & 4.5\% \\
 & D2 & 647 & 594&  423 & 232 & 58 & 56 & 292 & 91.6\% & 2.3\% & 5.8\% \\
 & D3 & 542 & 521& 357 & 209 & 43 & 39 & 289 & 86.3\% & 5.0\% & 8.7\% \\
 & D4 & 436 & 460& 304 & 186 & 30 & 24 & 289 & 75.3\% & 12.5\% & 12.2\% \\
 & D5 & 325 & 415& 273 & 161 & 22 & 12 & 290 & 51.6\% & 32.6\% & 15.8\% \\
\cline{1-12}
\end{tabular}
\end{center}}
{\caption{\label{tbl:masslines}\footnotesize{\bf Some Key Masses for
Model Lines of Table~4}. 
%\textbf{NOTE: For some reason JHEP style does not like references in Table captions, so this ``5'' is hard entered!} 
Low-lying superpartner masses are given in units of GeV as
well as the wavefunction composition of the LSP neutralino. Here $m_{\tilde{q}_1}$ is the lightest first generation squark, $m_{\tilde{t}_1}$ is the lighter stop, and we have
defined the two mass differences $\Delta^{\pm} \equiv
m_{\wtd{C}_1^{\pm}} - m_{\wtd{N}_1^{0}}$ and $\Delta^{0} \equiv
m_{\wtd{N}_2^{0}} - m_{\wtd{N}_1^{0}}$.}}
\end{table}
%=================================================================

%=================================================================

From Eq.~(\ref{gaugino1}), we see that the
magnitude of all three soft gaugino mass parameters will diminish as
$\alpha_g$ is varied from $\order(1)$ negative values to $\order(1)$
positive values. The effect is strongest for the gluino,
as the contribution to its mass from gauge mediation is proportional to the value of $g_3^2(M_{\rm mess})$
at the messenger scale. As mentioned in Sec.~\ref{sec:scan}, we  expect some value of
$\alpha_g$ to exist above which the gluino will become the LSP, which happens for $\alpha_g = 0.5$ and~$1.0$ in model line~B.  However, states which are charged only under $U(1)_Y$, such
as the right-handed sleptons, are largely unaffected by the
variation in $\alpha_g$ since the threshold correction to their soft
masses scales as $g_1^4(M_{\rm mess})$. As a result the lightest
stau will have a roughly constant mass across the entire model line.  
For $\alpha_g < 0$, there can be points for which the lightest neutralino is heavier than the 
lightest stau, which occurred for $\alpha_g = -1.0$ in model
line~A. The other points all yielded a reasonable spectrum and proper
electroweak symmetry breaking at the low-energy
scale.

The collider phenomenology of these models is dictated first and
foremost by the overall mass scale of the superpartners --
particularly those which carry $SU(3)$ quantum numbers. The masses
of these states vary dramatically with $\alpha_g$. The mass of the
gluino and lightest stop are listed in the first two columns of
Table~\ref{tbl:masslines}. Model lines~C and~D were chosen to
produce very light squarks and gluinos.  While such deflected mirage mediation models typically run into challenges with the dark matter and Higgs bounds, they were chosen here as examples for comparison because they are much more favorable with respect to the LHC than cases with heavier superpartners.
Model lines A~-~C all involve a mass
for the lightest Higgs state which satisfies $m_h \geq 113\GeV$ over
all $\alpha_g$ values. For line~D, we have $m_h
\leq 112\GeV$ along the model line. However, we will retain this
model line as the signatures are
representative of a large class of deflected mirage mediation models. 

As has been pointed out recently~\cite{Feldman:2008hs}, once event
rates are normalized to the overall mass scale of the colored
superpartners the next most important factor determining the
inclusive signatures for a model at the LHC is the hierarchy of
low-lying superpartner masses. This is particularly true for
leptonic signatures produced through the production and decay of
light neutralino and chargino states. A comprehensive examination of
the possible hierarchy patterns in deflected mirage mediation
models in the manner of \cite{Feldman:2007zn,Berger:2008cq} is
beyond the scope of the present study and will be presented
elsewhere \cite{dmmhierarchies}.
Here we will simply list the value of the lightest neutralino mass,
the lightest slepton mass (generally a scalar tau), the gluino and
lightest stop mass, and the two mass differences between the
lightest neutralino and the next two lightest gaugino states which
we denote as
\begin{eqnarray} \Delta^{0} &\equiv&
m_{\tilde{N}_2^{0}} - m_{\tilde{N}_1^{0}}, \label{deltazero} \\
\Delta^{\pm} &\equiv& m_{\tilde{C}_1^{\pm}} - m_{\tilde{N}_1^{0}}.
\label{deltaplus}
\end{eqnarray}
These values are collected in Table~\ref{tbl:masslines}. The mass differences $\Delta^{\pm}$ and $\Delta^0$ decrease monotonically with increasing $\alpha_g$ values because the gaugino spectrum becomes more squeezed, with the result that eventually the wino is lighter than the bino.  
This implies a softening of the leptonic decay products of cascade decays involving
these states. Hence, one typically encounters a point at which the on-shell decays of the chargino (or
second-lightest neutralino) to a slepton become kinematically
forbidden -- further suppressing leptonic final state signatures.
%========= table of key masses ==========
\begin{table}[t]
{\begin{center}
\begin{tabular}{|ll||c|c||c|c|c|c|}
\multicolumn{2}{c}{Model Point} &
\multicolumn{1}{c}{$\sigma_{\SUSY}$ (pb)} &
\multicolumn{1}{c}{Trigger Eff.} & \multicolumn{1}{c}{Multijet} &
\multicolumn{1}{c}{1~Lepton} &
\multicolumn{1}{c}{OS~Dilepton} & \multicolumn{1}{c}{Trilepton}\\
\cline{1-8}
\multicolumn{2}{l}{Line A} & \multicolumn{6}{c}{} \\
\cline{1-8}
 & A2 & $1\times10^{-3}$ & 98.8\% & 7794 & 3846 & 687 & 213 \\
 & A3 & $5\times10^{-3}$ & 99.1\% & 8238 & 3741 & 360 & 105 \\
 & A4 & 0.02 & 98.4\% & 6171 & 5976 & 823 & 252 \\
 & A5 & 0.21 & 73.8\% & 1447 & 31 & 3 & 2 \\
\cline{1-8}
\multicolumn{2}{l}{Line B} & \multicolumn{6}{c}{} \\
\cline{1-8}
 & B1 & 0.38 & 98.4\% & 4339 & 4031 & 1486 & 447 \\
 & B2 & 1.54 & 96.8\% & 3155 & 3441 & 379 & 75 \\
 & B3 & 5.56 & 88.0\% & 2409 & 182 & 0 & 0 \\
\cline{1-8}
\multicolumn{2}{l}{Line C} & \multicolumn{6}{c}{} \\
\cline{1-8}
 & C1 & 0.25 & 98.9\% & 8798 & 3784 & 398 & 90 \\
 & C2 & 0.59 & 98.6\% & 7932 & 3588 & 310 & 68 \\
 & C3 & 1.45 & 98.0\% & 5591 & 3718 & 499 & 102 \\
 & C4 & 3.80 & 96.1\% & 2931 & 3577 & 353 & 76 \\
 & C5 & 11.71 & 90.2\% & 2785 & 871 & 12 & 2 \\
\cline{1-8}
\multicolumn{2}{l}{Line D} & \multicolumn{6}{c}{} \\
\cline{1-8}
 & D1 & 12.7 & 95.9\% & 2680 & 2728 & 654 & 145 \\
 & D2 & 27.0 & 94.0\% & 2274 & 2195 & 309 & 48 \\
 & D3 & 61.1 & 91.0\% & 1328 & 1278 & 132 & 16 \\
 & D4 & 152.0 & 84.6\% & 759 & 660 & 34 & 2 \\
 & D5 & 459.7 & 67.2\% & 365 & 109 & 4 & 1 \\
\cline{1-8}
\end{tabular}
\end{center}}
{\caption{\label{tbl:xseclines}\footnotesize{\bf Gross LHC Features
for Model Lines of Table~4}. 
%\textbf{NOTE: For some reason JHEP style does not like references in Table captions, so this ``5'' is hard entered!} 
The total cross section for production of
superpartners and {\tt PGS4} level one trigger efficiency are given
in the first two columns. The following four columns give the number
of events in each channel per 50,000 generated events. The
definitions of these signatures are modified slightly from those of
Section~4.1. 
%\textbf{Again, hard entered reference!}
}}
\end{table}
%=================================================================
The above properties are common to many models in which
anomaly-mediated supersymmetry breaking becomes
important~\cite{Gherghetta:1999sw}. Indeed, as $\alpha_g$ is
increased many properties of the gaugino sector are more 
and more like the anomaly mediation limit, since the moduli/gravity mediated contributions cancel with the gauge mediated contributions in this case.   As a partial illustration of this
trend, we give the wave-function composition of the lightest
neutralino in terms of bino, wino and Higgsino percentages in the
final three columns of Table~\ref{tbl:masslines}. While cases such
as model line~C will exhibit the bino-like LSP behavior
characteristic of  mSUGRA/CMSSM models, others such as lines~B
and~D will allow for a much richer set of LSP properties, including
cases with a ``well-tempered'' neutralino~\cite{ArkaniHamed:2006mb}.

%%%%%%%%%%%%%%%%%%%%

To analyze the signatures of these models at the LHC, 50,000~events
were generated for each of the points in Table~\ref{tbl:lines} at
$\sqrt{s}=14$~TeV using {\tt PYTHIA 6.4}. The decision to use a
fixed number of events, as opposed to a certain fixed integrated
luminosity, is based on the widely differing total cross-sections
for supersymmetric particle production across these model lines. The
total supersymmetric cross-sections range from an exceedingly small
1~fb for Point~A2 to the much larger value of 0.46~nb for Point~D5, roughly
following the mass scale of the $SU(3)$-charged superpartners.
The overall supersymmetric cross-sections and combined level one
trigger efficiencies are collected in Table~\ref{tbl:xseclines}. As
mentioned in the previous subsection, the drop in trigger
efficiencies with increasing $\alpha_g$ is due in part to the
diminishing mass gaps between low-lying gaugino states, resulting in
softer jets and leptonic decay products and fewer events entering the sample
via leptonic triggers.

While the total event rate for a fixed integrated luminosity clearly
distinguishes the various points in each line, we are here
interested in how the introduction of a non-vanishing $\alpha_g$
value changes the collider phenomenology from the pure mirage
mediation case with $\alpha_g = 0$. In order to make meaningful
statements, therefore, we will work with our fixed event rate
samples.   For simplicity and clarity of presentation, in the analysis that follows  we consider the deflected mirage mediation signal only and do not include Standard Model
backgrounds.   All signature selection criteria begin with a
cut on missing transverse energy of $\met \geq 100 \GeV$ and a cut
on transverse sphericity given by $S_T \geq 0.2$. Additional
selection cuts are imposed as described below.

%=(1)=============== plots of MET (A and D) =====================
\begin{figure}[t]
\begin{center}
    {
      \includegraphics[scale=0.6]{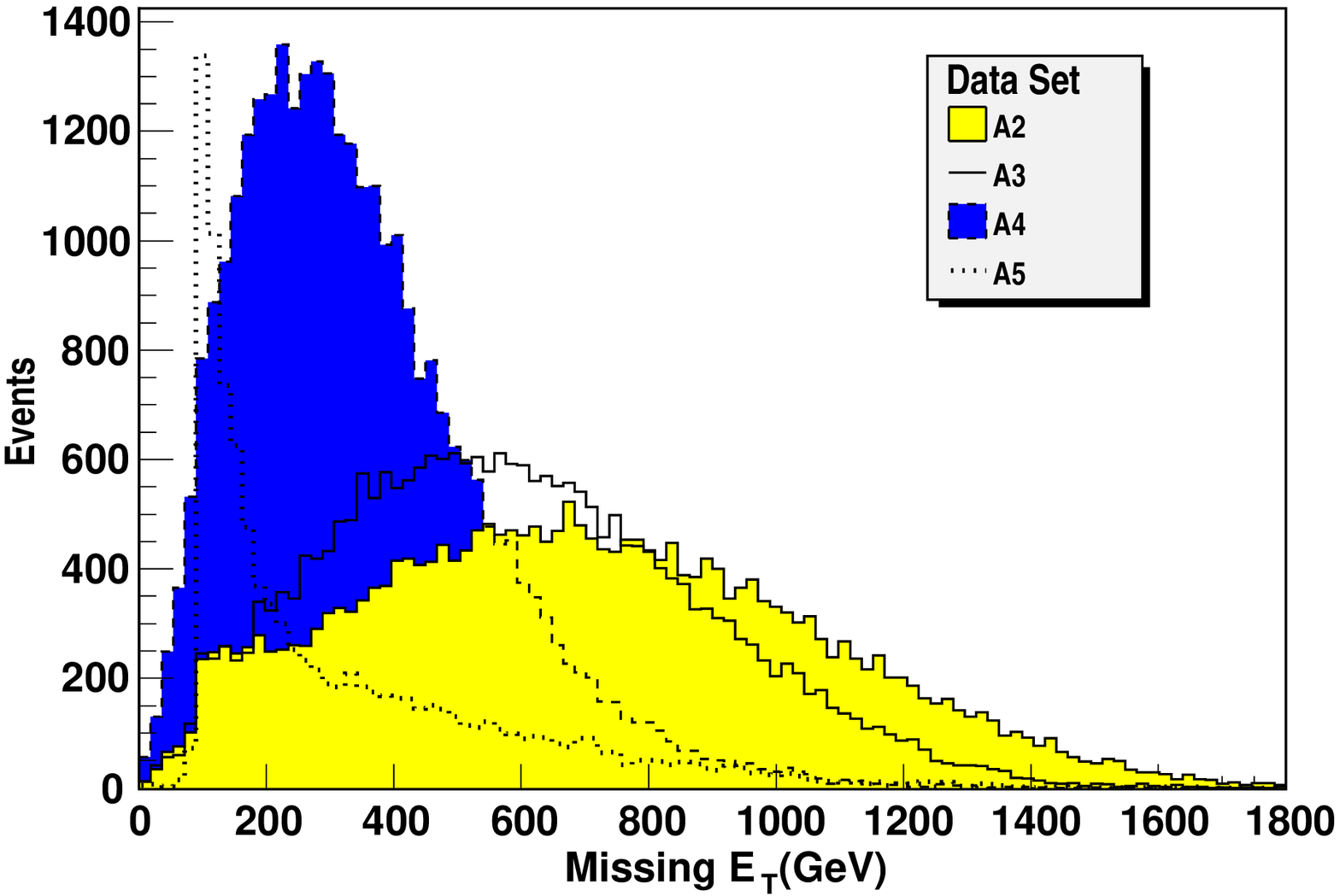}
    }
    {
      \includegraphics[scale=0.6]{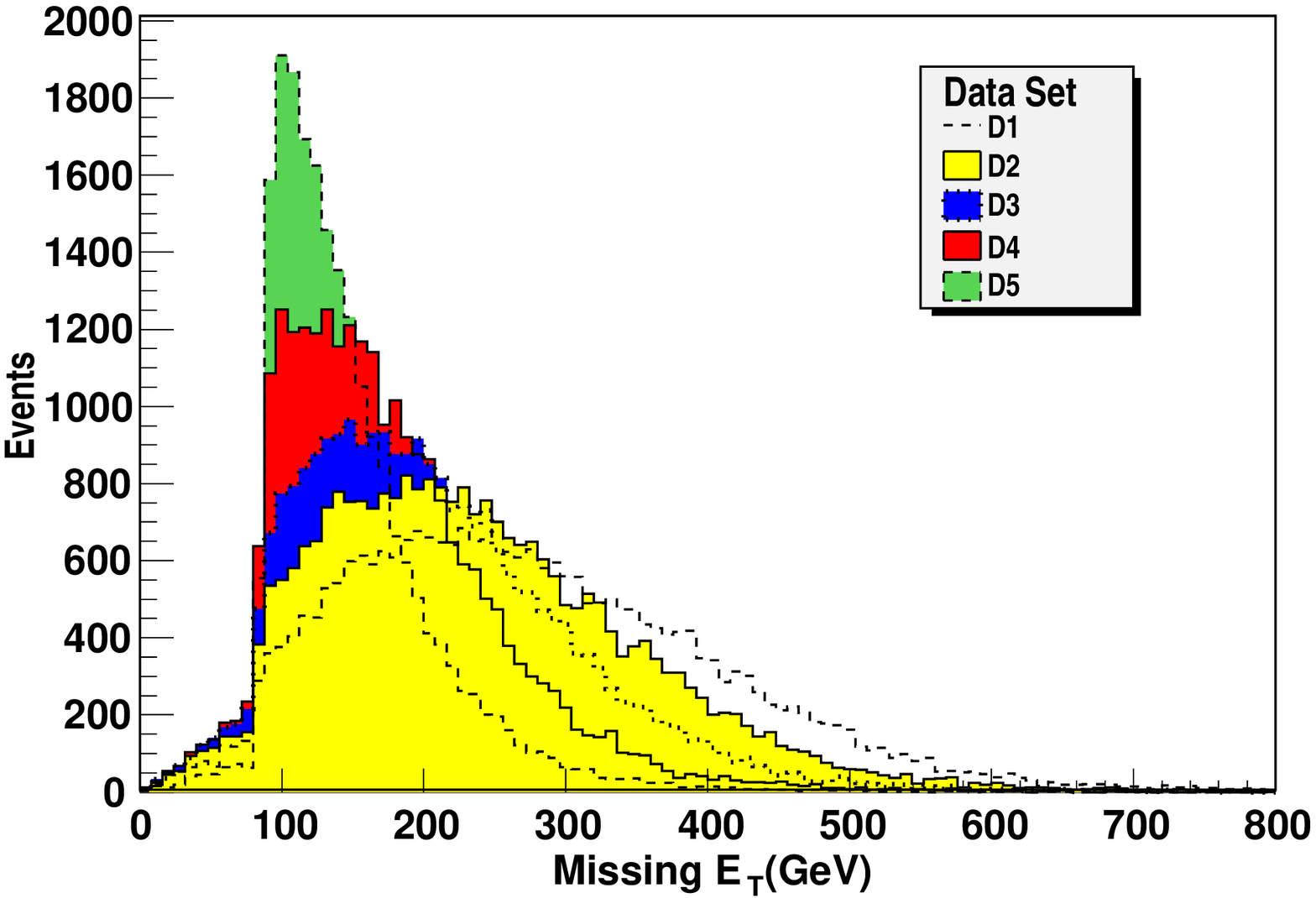}
    }
\caption{\label{plot:MET}\footnotesize{\textbf{Missing Transverse
Energy Distribution for Model Lines~A and~D.} Upper plot is for
line~A while lower plot is for line~D. Note that the {\tt PGS4}
level one trigger menu includes a selection on events with $\met
\geq 90\GeV$, as indicated by the sharp change in the distributions
for line~D.}}
\end{center}
\end{figure}
%=================================================================

We begin with the inclusive multijet signature~\cite{Baer:1995nq}
which is favored as a supersymmetry discovery channel. In this
subsection we will define this channel by the requirement of at
least 4~jets, the two hardest of which satisfy $p_T^{\rm jet} \geq
150 \GeV$ while the third and fourth hardest must satisfy $p_T^{\rm
jet} \geq 50 \GeV$. We also impose a veto on leptons ($e$ and $\mu$)
and require $\met \geq 200 \GeV$ for this channel. The total number
of such events for each model point is listed under the heading
``Multijet'' in Table~\ref{tbl:xseclines}. The dramatic fall in the
event count for larger values of $\alpha_g$ is in part due to the
rather severe cut on missing energy. In Figure~\ref{plot:MET} we
plot the distribution in $\met$ across all events satisfying $S_T
\geq 0.2$ for lines~A and~D. The {\tt PGS4} level one trigger
requires $\met \geq 90\GeV$ for the inclusive $\met$ trigger,
producing a sharp drop in the observed event rate below this
threshold. In the region $\alpha_g \gappeq 0.5$ the distribution is
clearly shifted to smaller values, with the majority of events
falling below the $\met = 200\GeV$ cut. Similar behavior occurs for
the other two model lines. The supersymmetric sample size can be
increased by relaxing this constraint, but only at the expense of
including more of the (already sizable) Standard Model background.

%=(2)=============== plots of Meff (A and C) =====================
\begin{figure}[t]
\begin{center}
    {
      \includegraphics[scale=0.6]{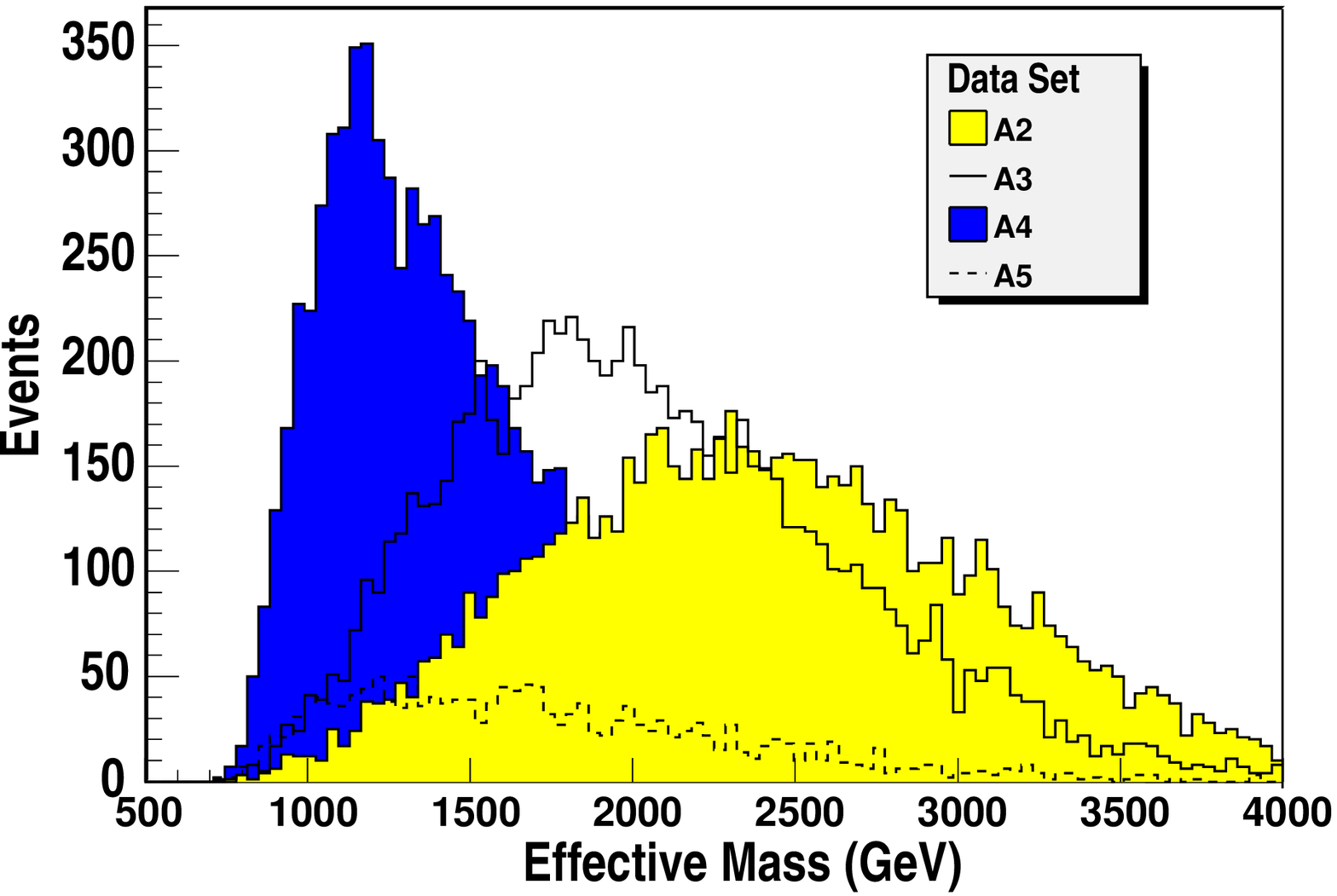}
    }
    {
      \includegraphics[scale=0.6]{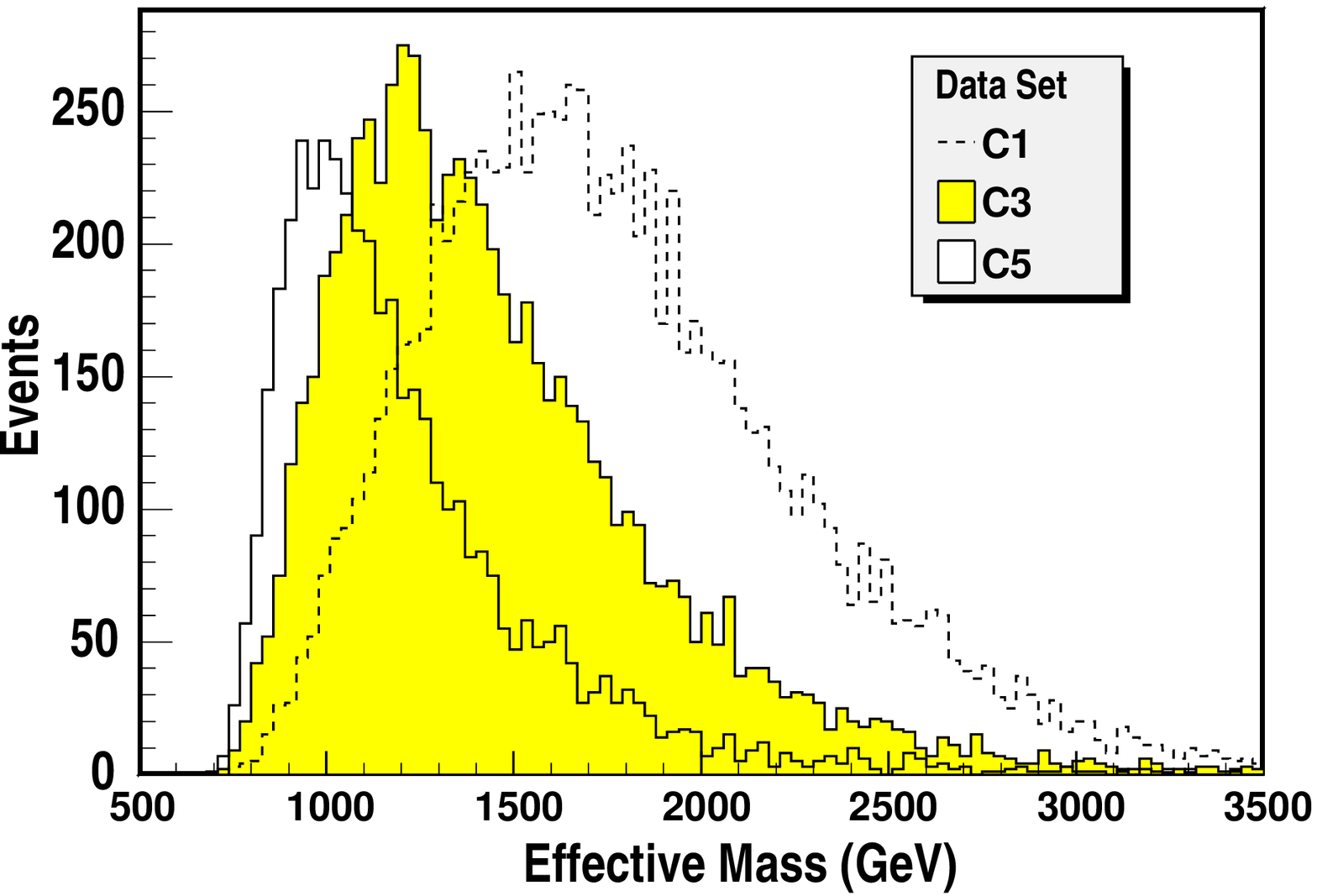}
    }
\caption{\label{plot:meff}\footnotesize{\textbf{Effective Mass
Distribution for Model Lines~A and~C.} The effective mass variable
is here defined as $M_{\rm eff} = \met + \sum_{i=1}^4 p^{\rm
jet_i}_T$. Smaller gluino masses are indicated by the shift in the
peak of this distribution to smaller values as $\alpha_g$ is varied
from $\alpha_g = -1$ to $\alpha_g = +1$.}}
\end{center}
\end{figure}
%=================================================================

In addition to being a discovery mode, the multijet channel has also
been suggested as a tool for crudely measuring the overall mass
scale of the superpartners. More specifically, the peak in the
distribution of the variable $M_{\rm eff}$, defined by the sum of
the missing transverse energy and the transverse momenta of the four
hardest jets in the event
\begin{equation} M_{\rm eff} = \met + \sum_{i=1}^4 p^{\rm
jet_i}_T\, , \label{Meff} \end{equation}
is known to track the mass of the lowest-lying colored superpartner,
most often the gluino~\cite{Hinchliffe:1996iu}. This continues to be
the case in deflected mirage mediation. In Figure~\ref{plot:meff} we
show the distribution in $M_{\rm eff}$ as defined by~(\ref{Meff})
for line~A and three values along line~C. The reduction of the
signal for Point~A5 is directly related to the decreased missing
energy in this case. For the other points in the two lines the total
number of events remains roughly constant, with the peak in the
distribution at a value roughly given by the gluino mass.

Events involving high~$p_T$ jets and isolated leptons with missing
transverse energy can also be excellent discovery modes for
supersymmetry~\cite{Baer:1991xs}. The final
three columns of Table~\ref{tbl:xseclines} list the number of events
involving one, two and three isolated leptons (respectively)
satisfying $p_T^{\ell} \geq 20\GeV$. Each of these signatures
requires two or more jets satisfying $p_T^{\rm jet} \geq 50 \GeV$
and $\met \geq 200 \GeV$. The dilepton signature involves precisely
two leptons of opposite sign, though they can be of any flavor (again, lepton here implies $e$ or $\mu$).

%=(3)=============== plots of pT lepton1 (c) =====================
\begin{figure}[t]
\begin{center}
    {
      \includegraphics[scale=0.6]{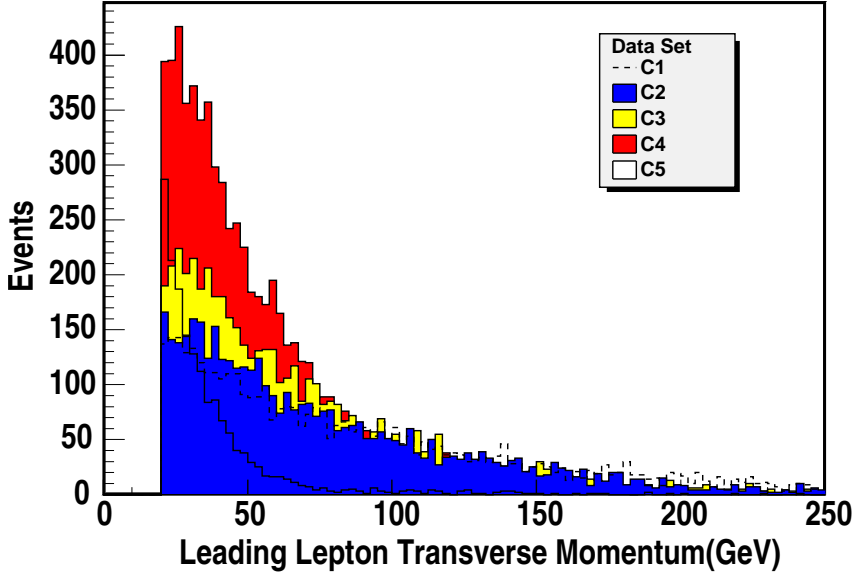}
    }
\caption{\label{plot:leppt}\footnotesize{\textbf{Transverse Momentum
of the Leading Lepton for Model Line C.} The transverse momentum of
the lepton with the largest $p_T$ value is given for the five models
of line~C. Distributions for the other model lines are similar.}}
\end{center}
\end{figure}
%=================================================================

Two features can be seen in the leptonic data of
Table~\ref{tbl:xseclines}. The broader feature is the general
reduction in leptonic activity as the value of $\alpha_g$ is
increased along each line. The reduction is most severe for the
multi-lepton signals when the mass gap between $\wtd{N}_2^0$ or
$\wtd{C}_1^{\pm}$ and the LSP drops below 50~GeV. For events with
one or more lepton, the $p_T$ of the leading lepton remains
relatively large. A representative example is given for the models
of line~C in Figure~\ref{plot:leppt}. The number of events with
$p_T^{\ell} \geq 100 \GeV$ remains relatively constant (hence the
roughly constant number of events in the ``1 Lepton'' channel),
while the number of events with softer leptons drops with increasing
$\alpha_g$. The softening of the second (or third) lepton in the
event is even more dramatic, resulting in fewer multi-lepton events.
For models with $\Delta^{\pm} \lappeq
5\GeV$, almost all signatures of the jets~+~leptons variety fail to
pass the trigger requirements or the leptonic $p_T$ cuts we impose.

The other feature involves the slight increase in multi-lepton
events between Points~A3 and~A4 in line~A and between Points~C2
and~C3 in line~C. Both are the result of the decreasing mass
differences between the lightest chargino/next-to-lightest
neutralino and the LSP. For Point~A3 the spoiler modes $\wtd{N}_2^0
\to \wtd{N}_1^0\,h$ and $\wtd{C}_1^{\pm} \to \wtd{N}_1^0\, W^{\pm}$
dominate their respective decay tables, with the former producing
mostly b-jets in the final state. For Point~A4, however, both of
these modes are kinematically inaccessible and the three body decay
modes $\wtd{N}_2^0 \to \wtd{N}_1^0\,f\bar{f}$ and $\wtd{C}_1^{\pm}
\to \wtd{N}_1^0\,f f'$ are populated. This results in a significant
increase in the number of multi-lepton final states. The effect is
also present in line~C, where the three body decay modes turn on
only for Points~C4 and~C5. For Point~C2 the dominant spoiler mode is
$\wtd{N}_2^0 \to \wtd{N}_1^0\,h$, but for Point~C3 this mode is
inaccessible and $\wtd{N}_2^0 \to \wtd{N}_1^0\,Z$ dominates, leading
to a dramatic increase in the number of opposite-sign (OS) dilepton
and trilepton final states.

%=(4)=============== plots of OSdilep (B and D) =====================
\begin{figure}[t]
\begin{center}
    {
      \includegraphics[scale=0.6]{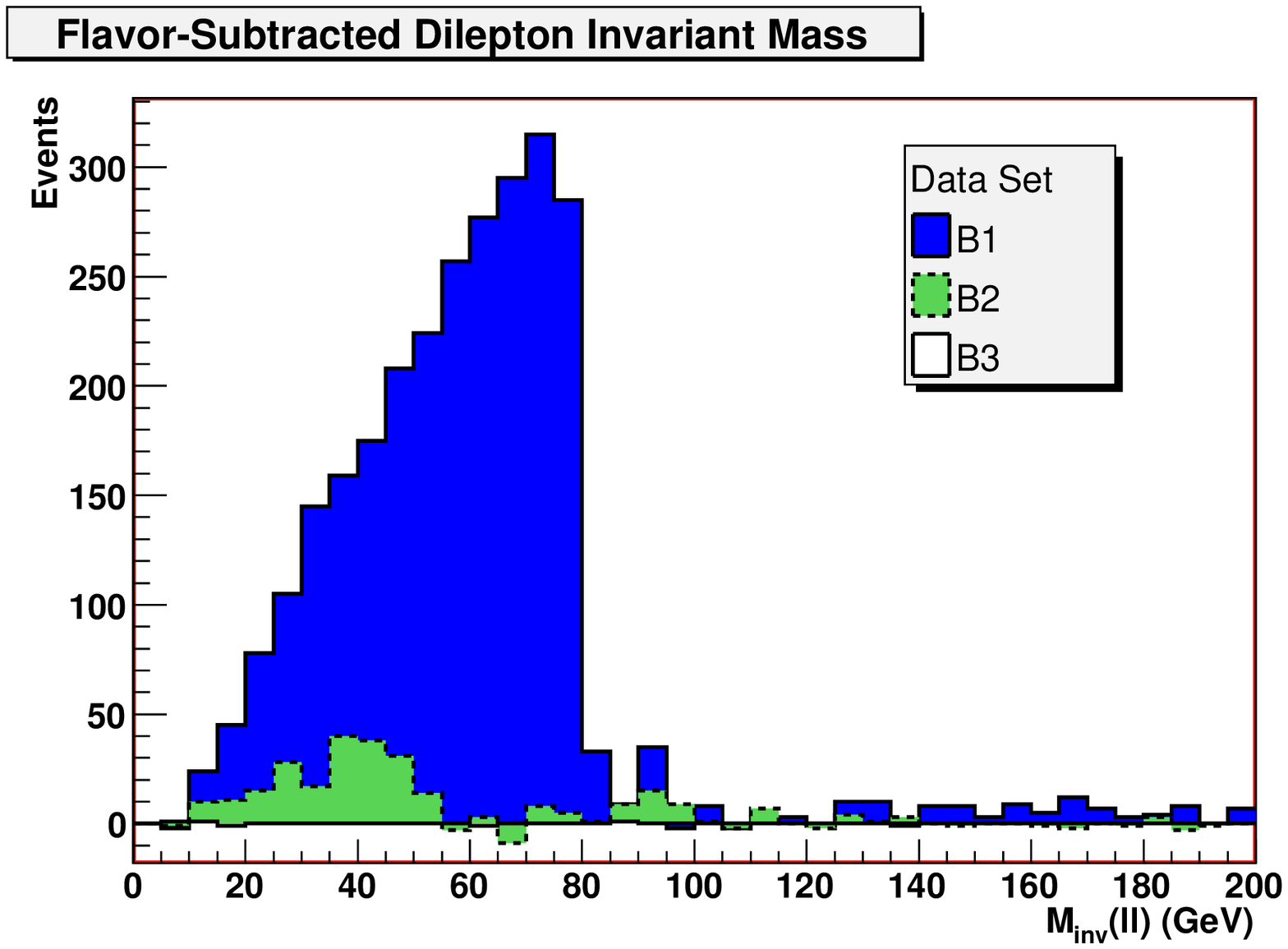}
    }
    {
      \includegraphics[scale=0.6]{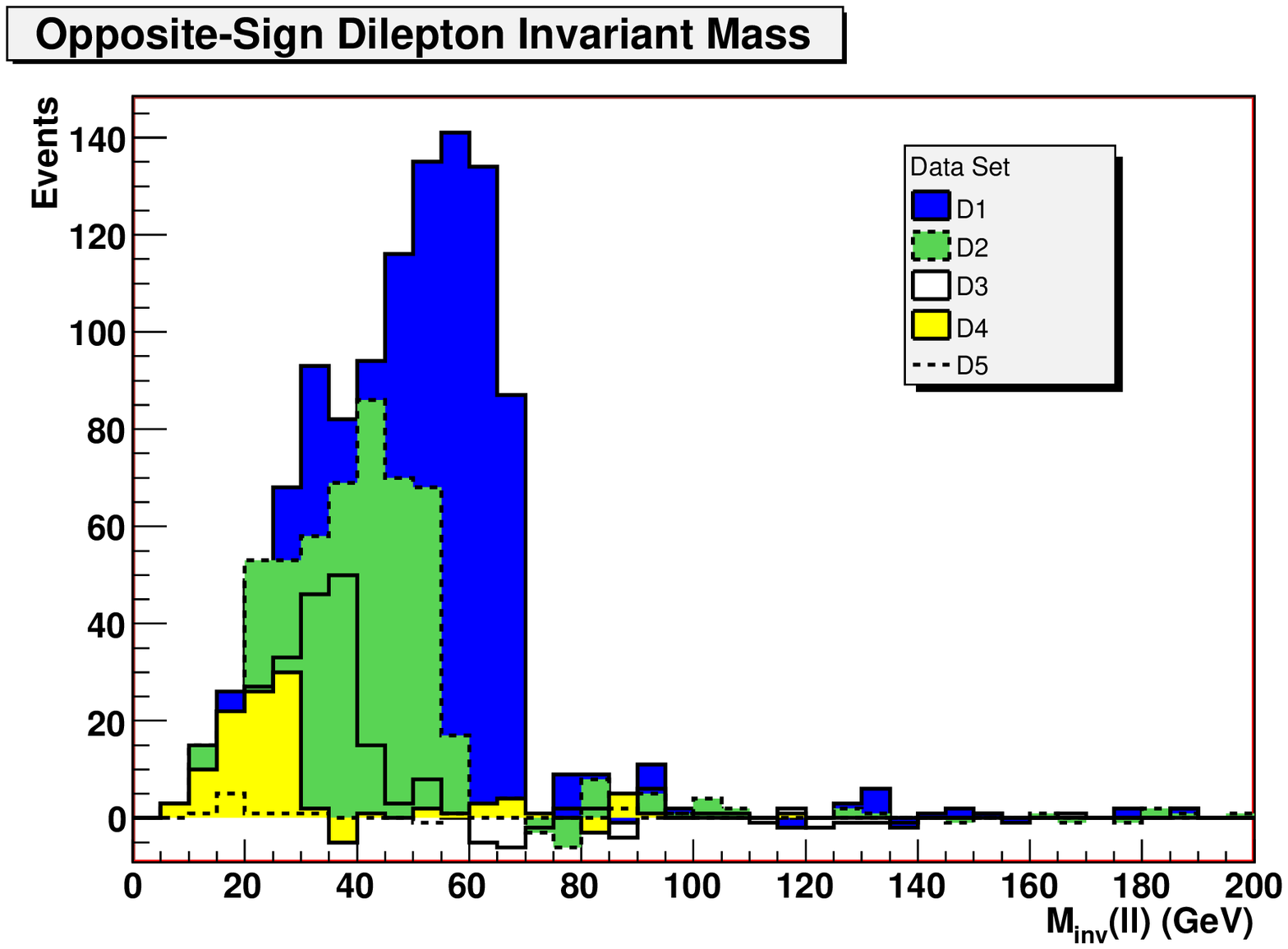}
    }
\caption{\label{plot:dilep}\footnotesize{\textbf{Flavor-Subtracted
Opposite-Sign Dilepton Invariant Mass Distribution for Model Lines~B
and~D.} The invariant mass distribution is formed from the subset
involving two leptons of opposite flavor which is subtracted from
that involving two of the same flavor ($e^+e^- +\mu^+\mu^- -e^+\mu^-
-e^-\mu^+$). For enough signal events, a
reasonable measurement of $\Delta^0$ is possible for both of these
model lines.}}
\end{center}
\end{figure}
%=================================================================

The opposite-sign dilepton channel is particularly important as it
provides crucial information on the mass differences between
low-lying gaugino eigenstates. For example, a typical strategy for
measuring the mass difference between light neutralinos is to form
the flavor-subtracted dilepton invariant mass for events with at
least two jets satisfying $p_T^{\rm jet} \geq 60$~GeV, at least
200~GeV of $\met$ and two opposite-sign
leptons~\cite{Hinchliffe:1996iu}. The invariant mass distribution
formed from the subset involving two leptons of opposite flavor is
subtracted from that involving two of the same flavor, i.e., the
combination ($e^+e^-+\mu^+\mu^--e^+\mu^--e^-\mu^+$), to reduce
background. For three-body decays involving $\wtd{N}_2^0 \to
\wtd{N}_1^0 \ell \bar{\ell}$ via a virtual slepton, an edge will
develop in the invariant mass distribution when $M_{\rm inv}=\Delta^0$, while for cascade decays involving on-shell sleptons, the edge appears at $M_{\rm inv}=\sqrt{(m_{\tilde{N}_2^0}^2-m_{\tilde{l}}^2)(m_{\tilde{l}}^2-m_{\tilde{N}_1^0}^2)/m_{\tilde{l}}^2}$.
This  distribution is plotted for lines~B and~D in
Figure~\ref{plot:dilep}. For both model lines, the first points (B1 and D1) are on-shell cascade decays, while the others are the three-body decays with off-shell sleptons.  For line~B, the third point does not yield sufficient events to produce a meaningful measurement.  However, point B1 clearly shows an edge in the invariant mass as expected.  Point B2, while having a much lower number of events, does indicate the expected shape based on the value of $\Delta^0$. For line~D an edge in the distribution can be clearly delimited in four of the five points.  For points D2--D4, the edges clearly track the
decreasing mass difference between the two lightest
neutralinos~(\ref{deltazero}) as $\alpha_g$ is increased.

%=(5)=============== plots of jet ratio (B and D) =====================
\begin{figure}[t]
\begin{center}
    {
      \includegraphics[scale=0.6]{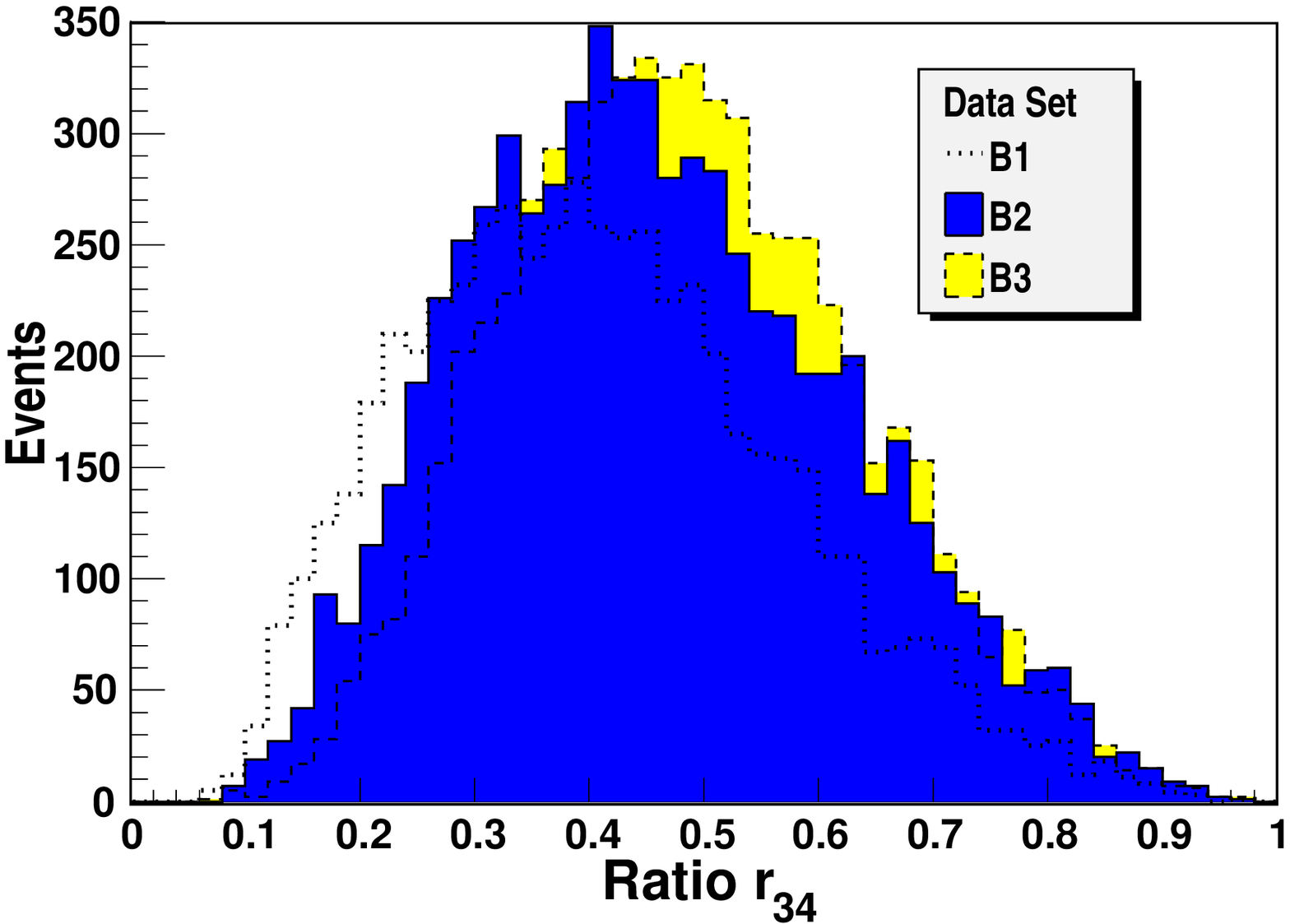}
    }
    {
      \includegraphics[scale=0.6]{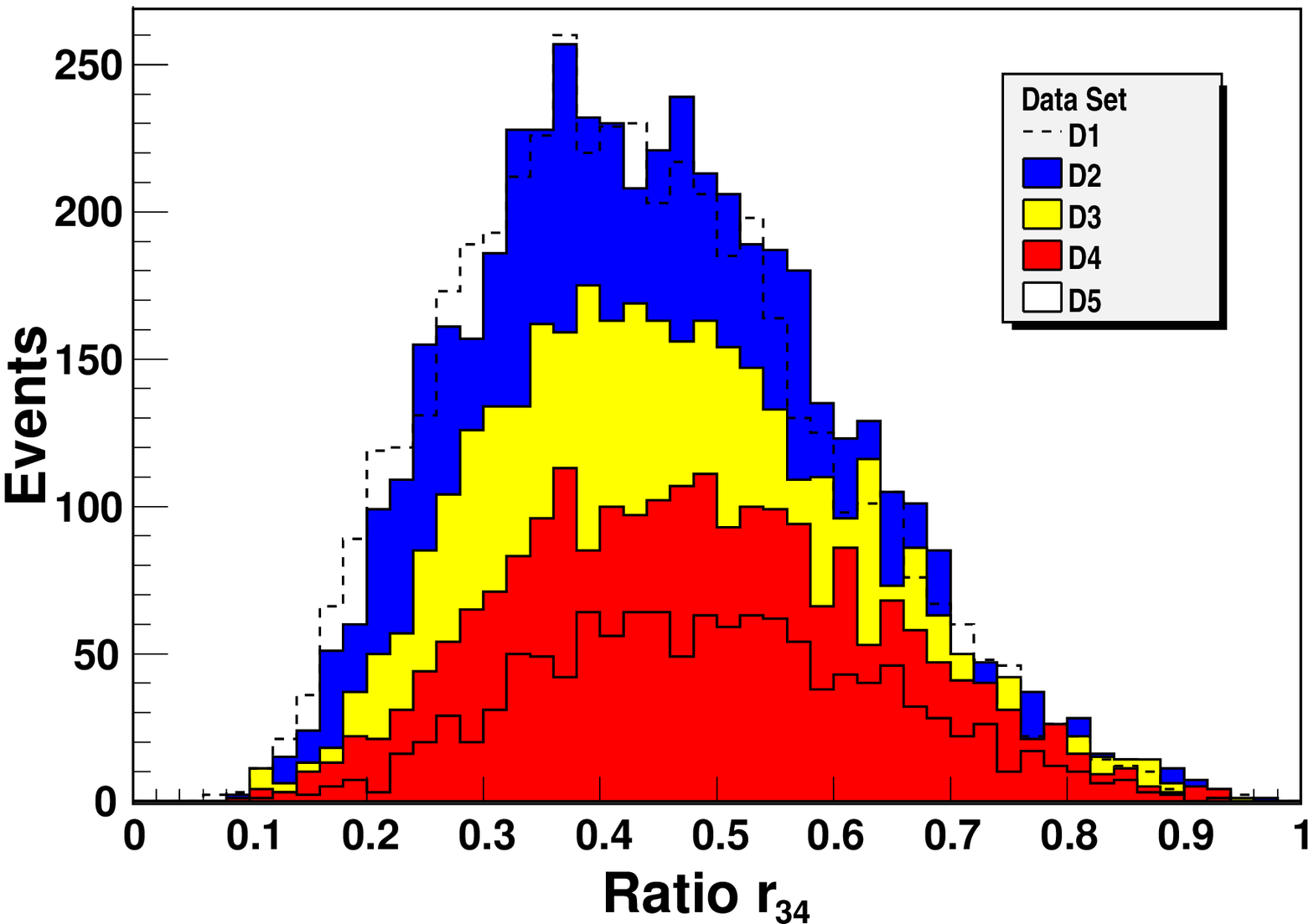}
    }
\caption{\label{plot:jetrat}\footnotesize{\textbf{Distribution of
Jet $p_T$ Ratio for Model Lines~B and~D.} The jet $p_T$
ratio $r_{\rm jet}=r_{34}$ is constructed from events with at least four jets
and is defined as $r_{\rm jet} = (p_T^{\rm jet 3} + p_T^{\rm jet
4})/(p_T^{\rm jet 1} + p_T^{\rm jet 2})$. }}
\end{center}
\end{figure}
%=================================================================

As discussed in Section~\ref{sec:scan}, for large thresholds ($\alpha_g>0$) the gauge mediated effects can produce a superpartner spectrum which is more compressed than the analogous case in pure mirage mediation. This can be seen in the
values of the invariant mass edge in the second plot in
Figure~\ref{plot:dilep}, and can also be seen
in cascade decays involving squarks and gluinos. For example, in
multijet events one can construct the following ratio
\begin{equation} r_{\rm jet} \equiv \frac{p_T^{\rm jet 3} + p_T^{\rm
jet 4}}{p_T^{\rm jet 1} + p_T^{\rm jet 2}}\, , \label{rjet}
\end{equation}
where $p_T^{\rm jet\,i}$ is the transverse momentum of the $i$-th
hardest jet in the event. For this signature we do not impose a
lepton veto, but we require that there be at least one jet with
$p_T^{\rm jet} \geq 100 \GeV$ and at least three more jets with
$p_T^{\rm jet} \geq 50 \GeV$. This signature was shown to be
effective in models based on the mirage pattern of gaugino
masses~\cite{Altunkaynak:2009tg}, and is here quite effective at
capturing the increasing softness of the products of cascade decays
as the value of $\alpha_g$ is changed. The distribution in the
values of this ratio is plotted in Figure~\ref{plot:jetrat} for
model lines~B and~D. The smaller the value of $r_{\rm jet}$, the larger
the mass difference between the initially produced gluino or squark
and the lighter neutralino and chargino states. As $\alpha_g \rightarrow 1$, the ratio tends towards $r_{\rm jet}
= 0.5$ in both cases. Plots for the other model lines give similar
results. Empirically we find the peak value of $r_{\rm jet}$ tracks the
value of $\alpha_g$ with roughly the same values across each of our
model lines. This suggest that signatures such as these may prove
extremely powerful at determining the value of $\alpha_g$ once
non-universality in the gaugino sector is firmly established.

%---------------------
\section{Conclusions}
\label{sec:conc}

We have investigated the collider phenomenology of deflected mirage mediation, a string-motivated ``mixed" supersymmetry breaking scenario in which effects from gravity mediation, anomaly mediation, and gauge mediation all contribute to the MSSM soft terms.  Our focus has been to compare the implications for LHC physics between deflected mirage mediation and the well-known mirage mediation framework, which includes gravity and anomaly mediation, but not gauge mediation.  The procedure was to explore deflected mirage mediation models together with standard mirage mediation benchmark models, either by directly comparing models with similar gaugino mass spectra, or by investigating the effects of switching on gauge mediation starting from pure mirage mediation scenarios.  

The results show that there is a broad variety of phenomenological outcomes within deflected mirage mediation, depending on the messenger scale and the size of the threshold effects from gauge mediation.   One interesting class of examples have a deflected gaugino mirage unification scale at TeV energies, leading to a squeezed spectrum in which the gluino can be the lightest colored superpartner, which in turn results in LHC signals with softer jets and leptons than in standard MSSM models.   The effects of gauge mediation can also have a large impact on the total superpartner production cross section, in some cases by several orders of magnitude.    For the deflected mirage mediation examples studied here, the most robust discovery mode will be the multijet channel.  The ratio of events with one lepton and high-$p_T$ jets to those with zero leptons is also capable of distinguishing between
the two different paradigms.

\acknowledgments
The work of B.A. and B.D.N. is supported by National Science Foundation Grant PHY-0653587.  L.E., I.W.K., and Y.R. are supported by the U.S. Department of Energy grant DE-FG-02-95ER40896.

\appendix

\section{Anomalous dimensions}
At one loop, the anomalous dimensions are given by 
\begin{eqnarray}
\gamma_i = 2 \sum_a g_a^2 c_a(\Phi_i) - \frac{1}{2}\sum_{lm} |y_{ilm}|^2,
\label{gammaexp}
\end{eqnarray}
in which $c_a$ is the quadratic Casimir, and $y_{ilm}$ are the normalized Yukawa couplings.  Here we will consider only the Yukawa couplings of the third generation $y_t$, $y_b$, and $y_\tau$.  For the MSSM fields $Q$, $U^c$, $D^c$, $L$, $E^c$, $H_u$ and $H_d$,
the anomalous dimensions are
\begin{eqnarray}
\gamma_{Q,i} &=& \frac{8}{3} g_3^2 + \frac{3}{2} g_2^2 + \frac{1}{30} g_1^2
- (y_t^2 + y_b^2) \delta_{i3}\nonumber \\
\gamma_{U,i} &=& \frac{8}{3} g_3^2 + \frac{8}{15} g_1^2
- 2 y_t^2 \delta_{i3},\;\; %\nonumber \\
\gamma_{D,i} =
%&=& 
\frac{8}{3} g_3^2 + \frac{2}{15} g_1^2
- 2 y_b^2 \delta_{i3},\nonumber \\
\gamma_{L,i} &=& \frac{3}{2} g_2^2 + \frac{3}{10} g_1^2
-y_\tau^2 \delta_{i3}, \;\; % \nonumber \\
\gamma_{E,i} =
%&=& 
\frac{6}{5} g_1^2
-2 y_\tau^2 \delta_{i3}, \nonumber \\
\gamma_{H_u} &=& \frac{3}{2} g_2^2 + \frac{3}{10} g_1^2
-3 y_t^2, \;\; 
%\nonumber \\
\gamma_{H_d} =
%&=& 
\frac{3}{2} g_2^2 + \frac{3}{10} g_1^2
- 3 y_b^2 - y_{\tau}^2,
\end{eqnarray}
respectively.
Above $M_{\rm mess}$, the beta function of the gauge couplings
changes because of the messenger fields.  However, $\gamma_i$ does not change according to
Eq.~(\ref{gammaexp}), and hence $\gamma'_i = \gamma_i$.  The $\dot{\gamma}_i$'s are given by the expression
\begin{eqnarray}
\dot{\gamma}_i=2\sum_a g_a^4b_a c_a(\Phi_i) - \sum_{lm} |y_{ilm}|^2b_{y_{ilm}},
\end{eqnarray}
in which $b_{y_{ilm}}$ is the beta function for the Yukawa coupling $y_{ilm}$.  The 
$\dot{\gamma}_i$'s are given by
\begin{eqnarray}
\dot\gamma_{Q,i}
&=& \frac{8}{3} b_3 g_3^4 + \frac{3}{2} b_2 g_2^4 + \frac{1}{30} b_1 g_1^4
- (y_t^2 b_t + y_b^2 b_b ) \delta_{i3} \nonumber \\
\dot\gamma_{U,i}
&=& \frac{8}{3} b_3 g_3^4 + \frac{8}{15} b_1 g_1^4
- 2 y^2_t b_t \delta_{i3}, \;\; % \nonumber \\
\dot\gamma_{D,i}=
%&=& 
\frac{8}{3} b_3 g_3^4 + \frac{2}{15} b_1 g_1^4
- 2 y^2_b b_b \delta_{i3} \nonumber \\
\dot\gamma_{L,i}
&=& \frac{3}{2} b_2 g_2^4 + \frac{3}{10} b_1 g_1^4
 - y_\tau^2 b_\tau \delta_{i3},\;\; 
 %\nonumber \\
\dot\gamma_{E,i}=
%&=& 
\frac{6}{5} b_1 g_1^4
 - 2 y_\tau^2 b_\tau \delta_{i3} \nonumber \\
\dot\gamma_{H_u}
&=& \frac{3}{2} b_2 g_2^4 + \frac{3}{10} b_1 g_1^4
 - 3 y^2_t b_t,\;\; % \nonumber \\
\dot\gamma_{H_d}
%&=&
=\frac{3}{2} b_2 g_2^4 + \frac{3}{10} b_1 g_1^4
 - 3 y_b^2 b_b - y^2_\tau b_\tau,  \label{dotgammaexp}
\end{eqnarray}
where
%\begin{eqnarray}
$b_t = 6 y_t^2 + y_b^2 -\frac{16}{3} g_3^2 - 3 g_2^2 - \frac{13}{15} g_1^2$, %\;\;
%\nonumber \\
$b_b =
%&=& 
y_t^2 + 6 y_b^2 + y_\tau^2 -\frac{16}{3} g_3^2 - 3 g_2^2
- \frac{7}{15} g_1^2$ %, \nonumber \\
and $b_\tau = 3 y_b^2 + 4 y_\tau^2 - 3 g_2^2 -\frac{9}{5} g_1^2$.
%\end{eqnarray}
$\dot\gamma^\prime_i$ is obtained by replacing $b_a$ with $b'_a = b_a + N$
in Eq.~(\ref{dotgammaexp}). 

Finally, ${\theta_i}$, which appears in the mixed modulus-anomaly term in the soft scalar mass-squared parameters, is given by
\begin{eqnarray}
\theta_i = 4 \sum_a g_a^2 c_a(Q_i) - \sum_{i,j,k} |y_{ijk}|^2
( 3- n_i -n_j- n_k).
\end{eqnarray}
For the MSSM fields, they take the form
\begin{eqnarray}
\theta_{Q,i} &=& \frac{16}{3} g_3^2 + 3 g_2^2 + \frac{1}{15} g_1^2
-2 ( y_t^2 (3-n_{H_u}-n_Q -n_U) + y_b^2 (3-n_{H_d} -n_Q - n_D )) \delta_{i3},
\nonumber \\
\theta_{U,i} &=& \frac{16}{3} g_3^2 + \frac{16}{15} g_1^2
- 4 y_t^2 (3-n_{H_u} - n_Q - n_U ) \delta_{i3}\nonumber \\
\theta_{D,i}&=& 
\frac{16}{3} g_3^2 + \frac{4}{15} g_1^2
- 4 y_b^2 (3- n_{H_d} - n_Q - n_D ) \delta_{i3}, \nonumber \\
\theta_{L,i} &=& 3 g_2^2 + \frac{3}{5} g_1^2
-2 y_\tau^2 ( 3- n_{H_d} - n_L - n_E ) \delta_{i3}\nonumber \\
\theta_{E,i} &=& 
\frac{12}{5} g_1^2
- 4 y_\tau^2 (3-n_{H_d} -n_L -n_E ) \delta_{i3}, \nonumber \\
\theta_{H_u} &=& 3 g_2^2 + \frac{3}{5} g_1^2
- 6 y_t^2 ( 3- n_{H_u} - n_Q - n_U )\nonumber \\
\theta_{H_d} &=& 
3 g_2^2 + \frac{3}{5} g_1^2
-6 y_b^2 ( 3- n_{H_d} - n_Q - n_D )
-2 y_\tau^2 ( 3- n_{H_d} - n_L - n_E ).
\end{eqnarray}
As in the case of $\gamma_i$, $\theta^\prime_i$ is the same as $\theta_i$.

%\clearpage
%%%%%%%%%%%%%%%%%%%%%%%%%%%%%%%%%%%%%%%%%%%%%%%%%%%%%%%%%%%%%%%%%%%%%%%%
%\pagebreak

\end{document}